\documentclass[prb,twocolumn,superscriptaddress,longbibliography]{revtex4-2}
\usepackage{amsmath,amssymb, mathrsfs}
\usepackage{stmaryrd}
\usepackage{latexsym}
\usepackage{graphicx}
\usepackage{indentfirst}
\usepackage{listings}
\usepackage{xcolor}
\usepackage{booktabs}
\usepackage{stmaryrd}
\usepackage{multirow}
\usepackage{verbatim}
\usepackage{makecell}
\usepackage{lineno,hyperref}
\usepackage{longtable}
\usepackage{hyperref}
\usepackage{upgreek}
\usepackage{bm}
\usepackage{bookmark}
\usepackage{enumitem}
\usepackage{comment}

\hypersetup{colorlinks,breaklinks,
            linkcolor=blue,urlcolor=blue,anchorcolor=blue,citecolor=blue}

\newcommand{\ud}{\mathrm{d}}
\newcommand{\txb}{\text{b}}

\newcommand{\txB}{\text{B}}
\newcommand{\txP}{\text{P}}
\newcommand{\txR}{\text{R}}
\newcommand{\txT}{\text{T}}

\newsavebox{\foobox}
\newcommand{\slantbox}[2][0]{\mbox{%
        \sbox{\foobox}{#2}%
        \hskip\wd\foobox
        \pdfsave
        \pdfsetmatrix{1 0 #1 1}%
        \llap{\usebox{\foobox}}%
        \pdfrestore
}}
\newcommand\unslant[2][-.25]{\slantbox[#1]{$#2$}}

\begin{document}

\title{%Dynamics of $\langle \mathbf{a}\rangle$ screw dislocations in $\alpha$-titanium: A multiscale approach or\\
Finite-Temperature Screw Dislocation Core Structures and Dynamics in $\alpha$-Titanium}

\author{Anwen Liu}
\affiliation{Department of Materials Science and Engineering, City University of Hong Kong, Hong Kong SAR, China}
\author{Tongqi Wen}
\affiliation{Department of Mechanical Engineering, The University of Hong Kong, Hong Kong SAR, China}
\author{Jian Han}
\email[]{jianhan@cityu.edu.hk}
\affiliation{Department of Materials Science and Engineering, City University of Hong Kong, Hong Kong SAR, China}
\author{David J. Srolovitz}
\email[]{srol@hku.hk}
\affiliation{Department of Mechanical Engineering, The University of Hong Kong, Hong Kong SAR, China}
%\email[Corresponding Authors: ]{srol@hku.hk, jianhan@cityu.edu.hk}

\begin{abstract}
A multiscale approach based on molecular dynamics (MD) and kinetic Monte Carlo (kMC) methods is developed to simulate the dynamics of an $\langle \mathbf{a} \rangle$ screw dislocation in $\alpha$-Ti. 
The free energy barriers for the core dissociation transitions and Peierls barriers for  dislocation glide as a function of temperature are extracted from the MD simulations (based on Machine Learning interatomic potentials and optimization); these form the input to kMC simulations.
Random walk dislocation trajectories from kMC agree well with those predicted by MD. 
On some planes, dislocations move via a locking-unlocking mechanism. 
Surprisingly, some dislocations glide in directions that are not parallel with the core dissociation direction. 
The MD/kMC multiscale method proposed is applicable to dislocation motion in simple and complex materials (not only screw dislocations in Ti) as a function of temperature and stress state. 

\end{abstract}

\maketitle

%%%%%%%%%%%%%%%%%%%%%%%%%%%%%%%%%%%%%%%%%%%%%%%%%%%%%%%%%%%%%%%%%%%%%%%%%%
\section{Introduction}

The major plastic deformation mechanism in crystalline metals is dislocation glide.  
%To understand and predict crystal plasticity, dislocation dynamics is a necessary and fundamental element. 
The motion of dislocations is (largely) controlled  by elastic stresses and intrinsic dislocation properties.  
%The driving force is the shear stress resolved on each slip plane, which can be easily analyzed by continuum elasticity. 
While these stresses are easily  analyzed in terms of continuum elasticity, dislocation motion  often occurs preferentially on planes other than those with the largest resolved shear stress.
This is associated with differences in glide resistance/lattice friction between different slip planes. 
This depends on the relative ease with which the dislocation core moves. 
The ease of glide, in turn, is sensitive to the dislocation core structure~\cite{Peierls_1940,Nabarro_1947,vitek1974theory,duesbery1991dislocation,vitek1992structure,cai2004dislocation}.
The dislocation core structure is, in many cases, temperature-dependent.
While knowledge of the core structure is essential, a quantitative link between core structure and dislocation dynamics is often elusive.
Here, we develop a multiscale approach to predict screw dislocation dynamics in $\alpha$-Ti.
%However, in this rationale, two issues are not yet handled carefully. 

%The first issue is how to acquire the accurate dislocation core structure at a finite temperature. 
Atomistic simulations are commonly employed to determine dislocation core structure, in part because the direct experimental determination of the structure is  demanding~\cite{yang2015imaging}. 
Since the atomic structure of the material is highly distorted with respect to that in a perfect crystal, quantum mechanical accuracy is often required to predict core structures; often achieved using density functional theory (DFT) calculations~\cite{clouet2009dislocation,romaner2010effect,Clouet2015locking,rodney2017ab}. 
Transition state theory-based methods (such as the nudged elastic band NEB method) are often employed to discern  the minimum energy path of a dislocation core as it traverses the slip plane~\cite{weinberger2013peierls,Clouet2015locking,rodney2017ab,kraych2019non}. 
While DFT methods are usually limited to ground-state (0~K) structures, finite-temperature \emph{ab initio} molecular dynamics are possible but too computationally costly for widespread use. 
Therefore, most finite-temperature dislocation core structure determination is based upon semi-empirical interatomic potential methods; e.g.,  Poschmann et al.~\cite{poschmann2022molecular}
studied the  core structure of an $\langle\mathbf{a}\rangle=a\langle10\overline{1}0\rangle$ screw dislocation in $\alpha$-Ti at finite temperature using a modified embedded atom method (MEAM) potential~\cite{hennig2008classical}.
Unfortunately, this MEAM potential fails to accurately reproduce all of the relevant 0~K  core structures and energies predicted by DFT~\cite{ghazisaeidi2012core}. 
Bond order potentials (BOPs)~\cite{aoki2007atom,cawkwell2005origin} were proposed to retain the quantum nature of atomic interactions in transition metals in a more cost-effective manner than DFT. 
However,  BOPs are both computationally costly and not easily implemented in molecular dynamics (MD) simulations~\cite{pettifor2000bounded}. 
The recently developed Deep Potentials (DPs)~\cite{han2017deep,wen2022deep} (a class of neural network potential) yield DFT accuracy with near empirical potential computational efficiency. 
Here, we employ the DP method to predict the structure of screw dislocation cores at  finite-temperature in $\alpha$-Ti.

The link between dislocation core structure and dislocation dynamics is often related to the assumption that the dislocation glide direction is consistent with the dislocation core dissociation direction. 
While such an assumption may be valid in some simple cases, its validity is far from assured in the case of more complex (non-cubic) materials, such as  hexagonal close packed (HCP) metals. 
Several  models  have been proposed to simulate  dislocation dynamics at the mesoscale, such as discrete dislocation dynamics (DDD)~\cite{po2014recent,bulatov2006computer,lesar2013computational} and kinetic Monte Carlo (kMC)~\cite{Lin1999kmc,Cai1999kmcSi,Cai2001kmc,Stukowski2015kmc,bulatov2006computer} methods. 
Extant mesoscale models do not explicitly incorporate the effects associated with the dislocation core structure. 
Here, we develop a mesoscale dislocation dynamics model that incorporates an explicit description of  the atomic-scale character of dislocation core structure. 

 While recent simulations (e.g., see~\cite{poschmann2022molecular}) focus on  long dislocation lines, here we focus on the intrinsic dislocation properties associated with the dislocation core structure. 
The admittedly important roles played by dislocation kinks in the motion of long dislocations are omitted here in order to provide a thorough examination of core effects without the complicated features of kink dynamics (which vary dramatically with, e.g., local dislocation curvature, junctions and interactions with other dislocations). 
The effects of core structure of short $\langle \mathbf{a}\rangle$ screw dislocation segments in $\alpha$-Ti based on the DP for Ti as a function of temperature are investigated.
Experimental observations~\cite{naka1988low} show that the edge dislocations are highly mobile and the yield strength of HCP Ti is governed by screw dislocation lattice friction. 
We report the results of MD simulations (based on machine learning potentials of quantum mechanical accuracy) of screw dislocation core structures, transitions between different core structures, statistical analysis of dislocation motion, and the determination of the kinetic parameters  (i.e., free energy barriers associated with migration, core structure transitions, ...) describing screw dislocation motion.  
We then perform kinetic Monte Carlo simulations of screw dislocation motion in Ti  incorporating these quantum mechanically accurate MD simulation parameters.
We examine the effect of both temperature and loading direction on dislocation core transitions and dislocation mobility. 
The results provide the basis for understanding non-Arrhenius screw dislocation mobility in metals with complex crystal structures. 

%%%%%%%%%%%%%%%%%%%%%%%%%%%%%%%%%%%%%%%%%%%%%%%%%%%%%%%%%%%%%%%%%%%%%%%%%%%%
\section{Atomistic dislocation structure and dynamics}

\subsection{Dislocation Core Structure}

%It has been long believed that how a dislocation moves is encoded in the atomic core structure of the dislocation~\cite{duesbery1991dislocation,cai2004dislocation}. 
%This belief motivates large amounts of efforts in acquiring accurate dislocation core structure by DFT calculations~\cite{clouet2009dislocation,rodney2017ab,Clouet2015locking}. 
%However, DFT calculations only predict the 0~K ground-state structure (ab-initio molecular dynamics is possible but impractically expensive). 

\begin{figure*}[t]
\includegraphics[width=0.98\linewidth]{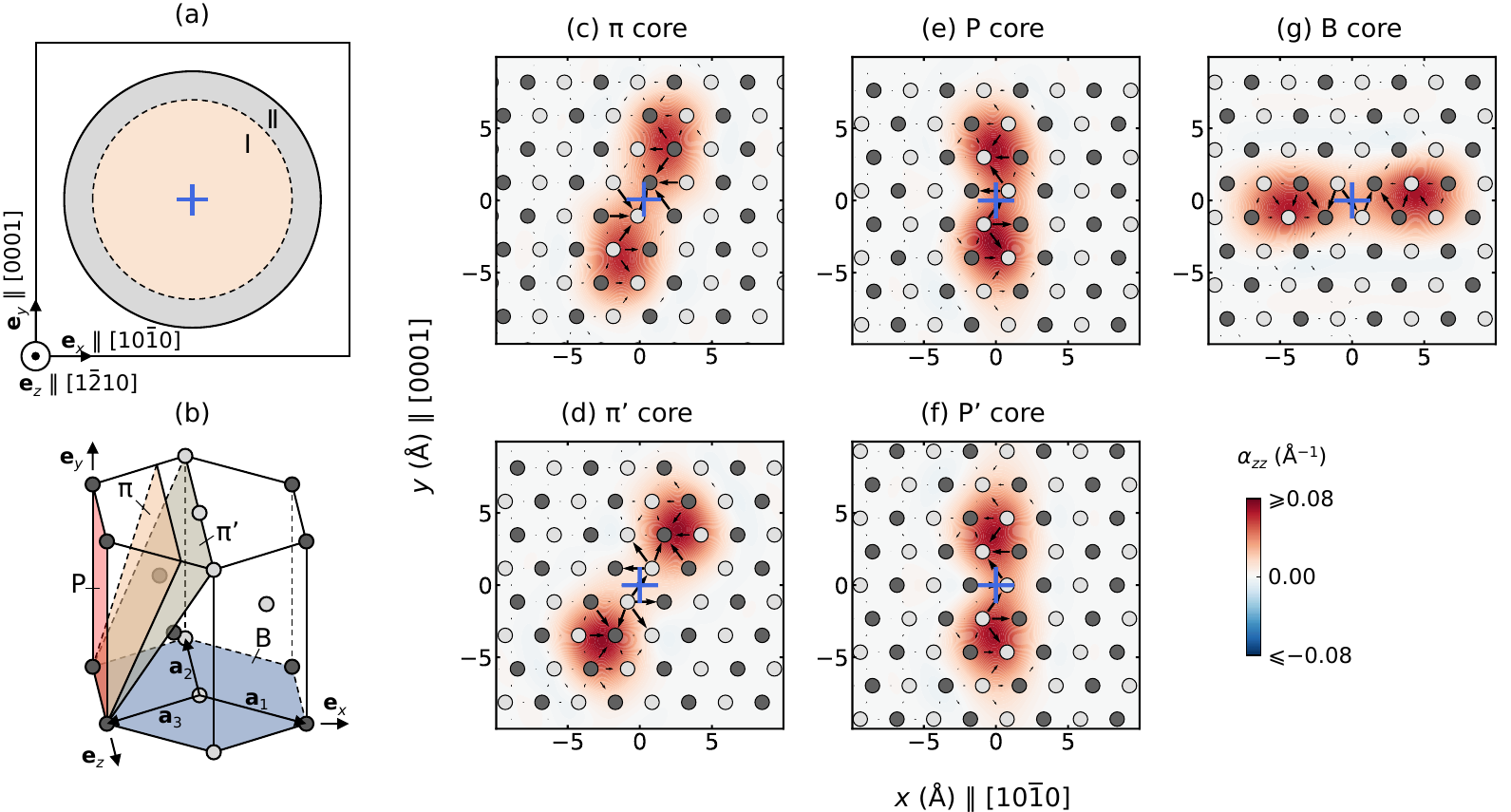}
\caption{\label{fig:model}
MD simulations of the $\langle\mathbf{a}\rangle$ dislocation core structure in $\alpha$-Ti. 
(a) The cylindrical model and  Cartesian coordinate system employed in the MD simulations. 
The blue ``$+$'' symbol indicates where the  screw dislocation was introduced. 
The atomic interaction in Region I and Region II are all modeled using DP, while atoms in Region II are treated via an Einstein model. 
(b) An HCP unit cell, where the P, $\unslant{\pi}$, $\unslant{\pi}'$ and B planes are identified. 
(c)-(g) The  dislocation cores observed in the MD simulations (following 10-20 energy minimization steps -- for improved visualization, see Method and Ref.~\cite{bulatov2006computer}).
The dark  and light gray circles denote  atoms on successive $(1\bar{2}10)$ planes (perpendicular to the $\mathbf{e}_z$-axis) in the perfect crystal. 
The arrows represent the differential displacement map while the color heat map shows the $\alpha_{zz}$ distribution. 
Blue ``$+$'' symbols indicate the core positions determined by the first moment of $\alpha_{zz}$. 
%The blue frame in (c) is a unit cell of the 2D lattice used in the kMC simulation; the edge lengths are $a_x = \sqrt{3}a/2$ and $a_y=c$.
}
\end{figure*}
We performed MD simulations to determine  $\langle \mathbf{a} \rangle$ screw dislocation core structure in $\alpha$-Ti; the simulation geometry is shown in Fig.~\ref{fig:model}a (see Methods).
By examining all the MD configurations from 300-900~K, we identified  five distinct core structures for the $\langle\mathbf{a}\rangle$ screw dislocation in HCP Ti. 
Figure.~\ref{fig:model}c-g show these five core structures with a differential displacement map~\cite{Vtek1970ddplot} (the black arrows) and the Nye tensor component $\alpha_{zz}$~\cite{nyetensor1,nyetensor2} (the contour). 
The $\alpha_{zz}$ map describes the screw component of the Burgers vector density. 
We find that the distribution of $\alpha_{zz}$ is highly delocalized in the form of a dipole, indicating that the core structure dissociates on a plane that includes   $[1\bar{2}10]$. 
For convenience, we denote the pyramidal plane by ``$\unslant{\pi}$'', the prismatic plane by ``P'' and the basal plane by ``B''. 
We find that two of the dislocation cores dissociate along the $\unslant{\pi}$ plane; i.e., the ``$\unslant{\pi}$ core'' (Fig.~\ref{fig:model}c) and ``$\unslant{\pi}'$ core'' (Fig.~\ref{fig:model}d). 
The dissociation plane of $\unslant{\pi}$ core is close to $(\bar{3}031)$ while that of $\unslant{\pi}'$ core is close to $(\bar{1}011)$, as shown in Fig.~\ref{fig:model}b. 
Two cores are dissociated along the P plane; i.e., the ``P core'' (Fig.~\ref{fig:model}e) and ``P$'$ core'' (Fig.~\ref{fig:model}f). 
The  $\alpha_{zz}(x,y)$ map for the P core possesses inversion symmetry roughly about the point $(0,0)$ while that for the P$'$ core possesses mirror symmetry roughly about the  $y=0$ line. 
We also identify a  dissociated core along the B plane; i.e., the ``B core'' (Fig.~\ref{fig:model}g).
The $\unslant{\pi}$, $\unslant{\pi}'$, P and P$'$ cores were found at all temperatures in our simulations. 
These three are consistent with the 0~K core structures predicted by  DFT calculations~\cite{Clouet2015locking}. 
The B core was observed only at high temperature, $T\gtrsim 400$~K.  

Since the $\unslant{\pi}$, $\unslant{\pi}'$, P and P$'$ cores are stable at 0~K, we can obtain their 0~K equilibrium structures by direct energy minimization based on different initial configurations (with the dislocation core centered at different positions). 
The $\unslant{\pi}$, $\unslant{\pi}'$, P and P$'$ core energies are $E_{\unslant{\pi}} = 544.8 \pm 0.43$~meV/{\AA}, $E_{\unslant{\pi}'} = 561.5 \pm 0.52$~meV/{\AA} and $E_{\txP} = E_{\txP'} = 547.4 \pm 0.34$~meV/{\AA}, respectively (see Supplementary Information, SI, for details).
The energy differences are around $E_{\txP}-E_{\unslant{\pi}} = 2.6$~meV/{\AA} and $E_{\unslant{\pi}'}-E_{\unslant{\pi}} = 16.7$~meV/{\AA}; for comparison, the DFT results~\cite{Clouet2015locking} are 5.7~meV/{\AA} and 11~meV/{\AA}, respectively. 
The dislocation core structures and energies at 0~K obtained from DP are reasonably consistent with  DFT results. 
Since the $\unslant{\pi}'$ core energy is much higher than the energies of the other cores and a nudged-elastic-band calculation~\cite{Clouet2015locking} shows that the $\unslant{\pi}'$ core energy is almost as high as the barrier for $\unslant{\pi}$ core glide, $\unslant{\pi}'$ core is not important for the thermodynamic and kinetic properties; hence, we  ignore the $\unslant{\pi}'$ core below. 
Since the P and P$'$ core energies are nearly equal, we do not distinguish the P and P$'$ cores below. 

%%%%
\begin{figure*}[tb]
\includegraphics[width=1\linewidth]{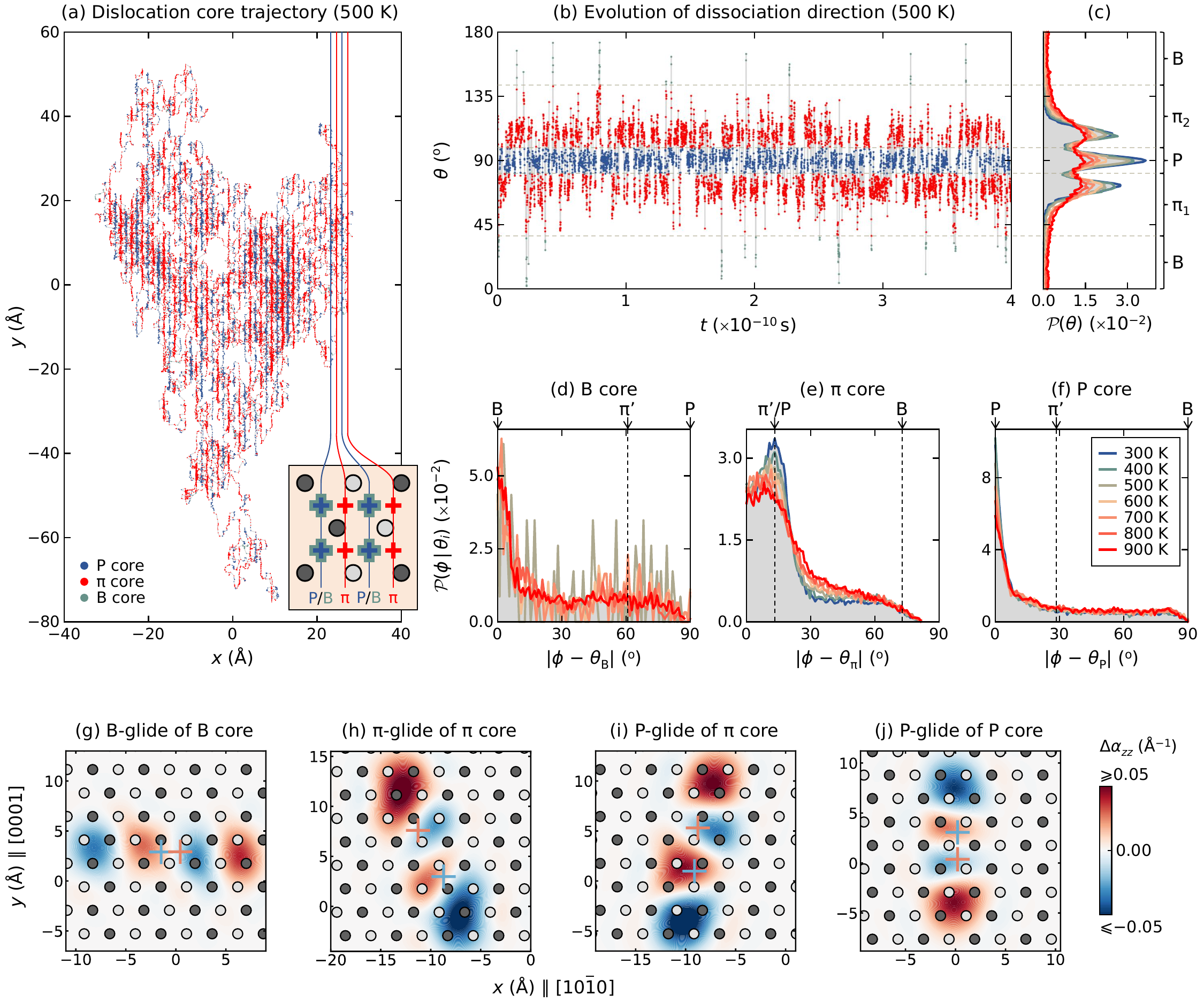}
\caption{\label{fig:set} 
Dislocation core trajectory, dissociation direction and glide direction. 
(a) The example trajectory of a dislocation core obtained by the MD simulation at $500$~K for 2~ns. 
The blue, red and green points denote the positions where the core structures are P, $\unslant{\pi}$ and B, respectively. 
The inset shows that the P/B cores and the $\unslant{\pi}$ cores should be distributed on alternating P planes. 
(b) Temporal evolution of the core dissociation direction $\theta$ extracted from the same MD simulation.
(c) Distribution of $\theta$ at the temperature ranging from 300~K to 900~K. 
The horizontal lines denote the boundaries we defined to distinguish the core structures at 500~K. 
(d)-(f) Distributions of the deviation of the glide direction $\phi$ away from the dissociation direction $\theta$ for the B, $\unslant{\pi}$ and P cores. 
(g)-(i) The differences of the $\alpha_{zz}$ maps at two moments, corresponding to the four dislocation core glide events observed in the simulation. 
The blue/red ``+'' symbol indicates the core position in the previous/next moment. 
}
\end{figure*}

\subsection{Dislocation Core Dynamics}

We focus on two aspects of dislocation core dynamics: core motion and core structure transitions. 
The former is analyzed based on the trajectory of the core position (Fig.~\ref{fig:set}a), while the latter requires recognition of instantaneous core structure. 
We automated the determination of the core position and the core dissociation direction based on the $\alpha_{zz}$ map. 
For convenience, we denote the $\alpha_{zz}$-weighted average of any quantity $A$ as
\begin{equation}
\bar{A}
= \frac{\iint A(x,y) \alpha_{zz}(x, y) \ud x \ud y}
{\iint \alpha_{zz}(x, y) \ud x \ud y},  
\end{equation}
Then, the core position is $\mathbf{r}_\text{core} \equiv \left(\bar{x}, \bar{y}\right)$; the core positions determined in this manner are indicated by the blue ``+''  in Figs.~\ref{fig:model}c-g. 
Figure~\ref{fig:set}a shows the core  trajectory at $500$~K. 
From the trajectory, we see that the dislocation random walk is anisotropic. 
The pattern is elongated along the $y$-axis, suggesting that the dislocation glides on the P plane most frequently. 

The glide direction as a function of time $\phi(t)$ is 
\begin{equation}
\phi(t)
= \arctan\left[
\frac{\bar{y}(t+\Delta t) - \bar{y}(t)}
{\bar{x}(t+\Delta t) - \bar{x}(t)}
\right], 
\end{equation}
where $\Delta t$ is the time step. 
The second moment of $\alpha_{zz}$ is the tensor $[C_{ij}]$, where $C_{11} = \overline{x^2}$, $C_{22} = \overline{y^2}$ and $C_{12} = C_{21} = \overline{xy}$. 
Suppose that the (normalized) eigenvector of $[C_{ij}]$ corresponding to the maximum eigenvalue is $\mathbf{e}^*$. 
Thus, the core dissociation direction at  $t$ is given by 
\begin{equation}\label{thetat}
\theta(t)
= \arctan\left[
{e_y^*(t)} \middle/ {e_x^*(t)}
\right]. 
\end{equation}

Figure~\ref{fig:set}b shows the temporal evolution of the dissociation direction $\theta$ at 500~K and Fig.~\ref{fig:set}c shows the distribution of the core dissociation direction $\theta$ at different temperatures. 
The three sharp peaks  at $\theta {\simeq}75^\circ$, $90^\circ$ and $105^\circ$ correspond to the $\unslant{\pi}_1$, P and $\unslant{\pi}_2$ cores ($\unslant{\pi}_1$ and $\unslant{\pi}_2$ are both $\unslant{\pi}$ cores). 
The peaks for the $\unslant{\pi}$ cores shift towards $\theta = 90^\circ$ as temperature increases because the $c/a$ ratio increases with temperature~\cite{wen_Ti}. 
%Note that ideally the angle between $\unslant{\pi}$ plane and B plane at 0~K is $\theta = \arctan(\sqrt{3}c/a) = 70^\circ$ while the $\unslant{\pi}$ core dissociation direction determined by $\alpha_{zz}$ (Eq.~\eqref{thetat}) is $75^\circ$. 
The peak for the $\unslant{\pi}'$ core, if it existed, would be  at $\sim 61^\circ$ and $\sim 110^\circ$; however, no peaks exist there, implying that the $\unslant{\pi}'$ core may be ignored. 
At a high temperature (e.g., $900$~K), there are shallow and broad humps  at $\theta = 0$ or $180^\circ$, indicating the existence of the B core at high temperature. 
The peaks,  signaling different cores,  broaden with increasing temperature. 
We set criteria to distinguish the core structures based on the distribution in Fig.~\ref{fig:set}c. 
We define the width of the orientation window for the P core as the minima between the P and $\unslant{\pi}$ cores; i.e.,  $\sim 90 \pm 9^\circ$ (the exact position of the  minima varies with temperature); see the two dashed lines near $\theta = 90^\circ$ in Fig.~\ref{fig:set}b or c. 
The boundary between the $\unslant{\pi}$ and B cores is not well defined. 
Since the probability for $\theta < 50^\circ$ or $> 130^\circ$ is almost zero, any quantity evaluated based on the core distribution is insensitive to the choice of this boundary. 
In practice, we choose the boundary between the $\unslant{\pi}$ and B cores at the $\theta$ in the middle between $0$ and the $\theta$ for the $\unslant{\pi}$ core (i.e., the $\theta$ for the second highest peak); see the two dashed lines close to $\theta = 90^\circ$ in Fig.~\ref{fig:set}b or c. 
Using this  criterion, we  can identify the core structure at any time during the MD simulation. 
We color each point on the trajectory (at $500$~K) shown in Fig.~\ref{fig:set}a according to the core structure at each time. 
We observe that  B cores (green points) are rare. 
The $\unslant{\pi}$ core (red points) and the P core (blue points) are largely distributed on  alternating P planes; this is consistent with the examination of the $\unslant{\pi}$ and P core positions in the Fig.~\ref{fig:set}a inset. 
The Fig.~\ref{fig:set}a inset shows that ideally the $\unslant{\pi}$ core is positioned between a dark gray P plane and a light gray P plane, while the P core and B core are  between two light gray or two dark gray P planes. 
Hence, the $\unslant{\pi}$ and P cores should be distributed on different P planes. 
Such consistency validates the  core structure recognition method. 

It is usually assumed that the dislocation core dissociation direction is the same as the dislocation glide direction; this need not be true. 
To investigate the correlation between the core dissociation direction ($\theta$) and the glide direction ($\phi$), we examine the  probability of different glide directions for particular core dissociations (characterized by the dissociation direction $\theta_i$ ($i \in \{\unslant{\pi}, \txP, \txB\}$). 
Figure~\ref{fig:set}d shows the glide direction distribution for the B core. 
Most frequently, the B core glides on the B plane (B-glide), corresponding to the peak at $|\theta - \phi_\txB|=0$. 
The B-glide of the B core is also seen in the difference between the $\alpha_{zz}$ maps at a pair of times: $\Delta\alpha_{zz} \equiv \alpha_{zz}(t_2) - \alpha_{zz}(t_1)$. 
As shown in Fig.~\ref{fig:set}g, the alternatively distributed negative and positive $\Delta\alpha_{zz}$ clouds lying on the B plane is a feature of B-glide of B core. 
The glide direction distribution for the $\unslant{\pi}$ core is shown  Fig.~\ref{fig:set}e.
This distribution exhibits a peak at $|\theta - \phi_{\unslant{\pi}}|\approx 15^\circ$ (at $T=300$~K), where $\theta_{\unslant{\pi}} \approx 75^\circ$. 
The angle between  the $\unslant{\pi}'$ plane (i.e., the $(\bar{1}011)$ plane) and the B plane at 300~K is $\phi = \arctan\left[(2/\sqrt{3})(c/a)\right] \approx 61^\circ$; so, if the $\unslant{\pi}$ core glides on the $\unslant{\pi}'$ plane ($\unslant{\pi}$-glide), $|\theta - \phi_{\unslant{\pi}}|\approx 14^\circ$. 
If the $\unslant{\pi}$ core glides on the P plane (P-glide), $|\theta - \phi_{\unslant{\pi}}| \approx 15^\circ$. 
This suggests the peak at ${\sim}15^\circ$ in Fig.~\ref{fig:set}e has contributions from both P-glide and $\unslant{\pi}$-glide. 
Careful examination, however, shows that  P-glide is much more frequent than $\unslant{\pi}$-glide. 
The $\unslant{\pi}$-glide and P-glide of $\unslant{\pi}$ core can be directly verified by the $\Delta\alpha_{zz}$ maps in Fig.~\ref{fig:set}h and i. 
The alternating negative and positive $\Delta\alpha_{zz}$ clouds on the $\unslant{\pi}$ plane is a feature of $\unslant{\pi}$-glide of the $\unslant{\pi}$ core while the off-line distribution of the negative and positive clouds is a feature of P-glide of the $\unslant{\pi}$ core. 
The observation that a $\unslant{\pi}$ core can glide on the P plane contradicts the assumption that the core glide and the core dissociation directions must be the same. 
Figure~\ref{fig:set}f shows the glide direction distribution for the P core. 
Clearly, the P core only glides on P plane. 
Again, the P-glide of the P core can be verified through considerations of the $\Delta\alpha_{zz}$ clouds distributed along the P plane, as shown in Fig.~\ref{fig:set}j. 

%%%%%%%%%%%%%%%%%%%%%%%%%%%%
\section{Model and parameterization of dislocation core dynamics}\label{sec:model}

\begin{figure*}[t]
\includegraphics[width=1\linewidth]{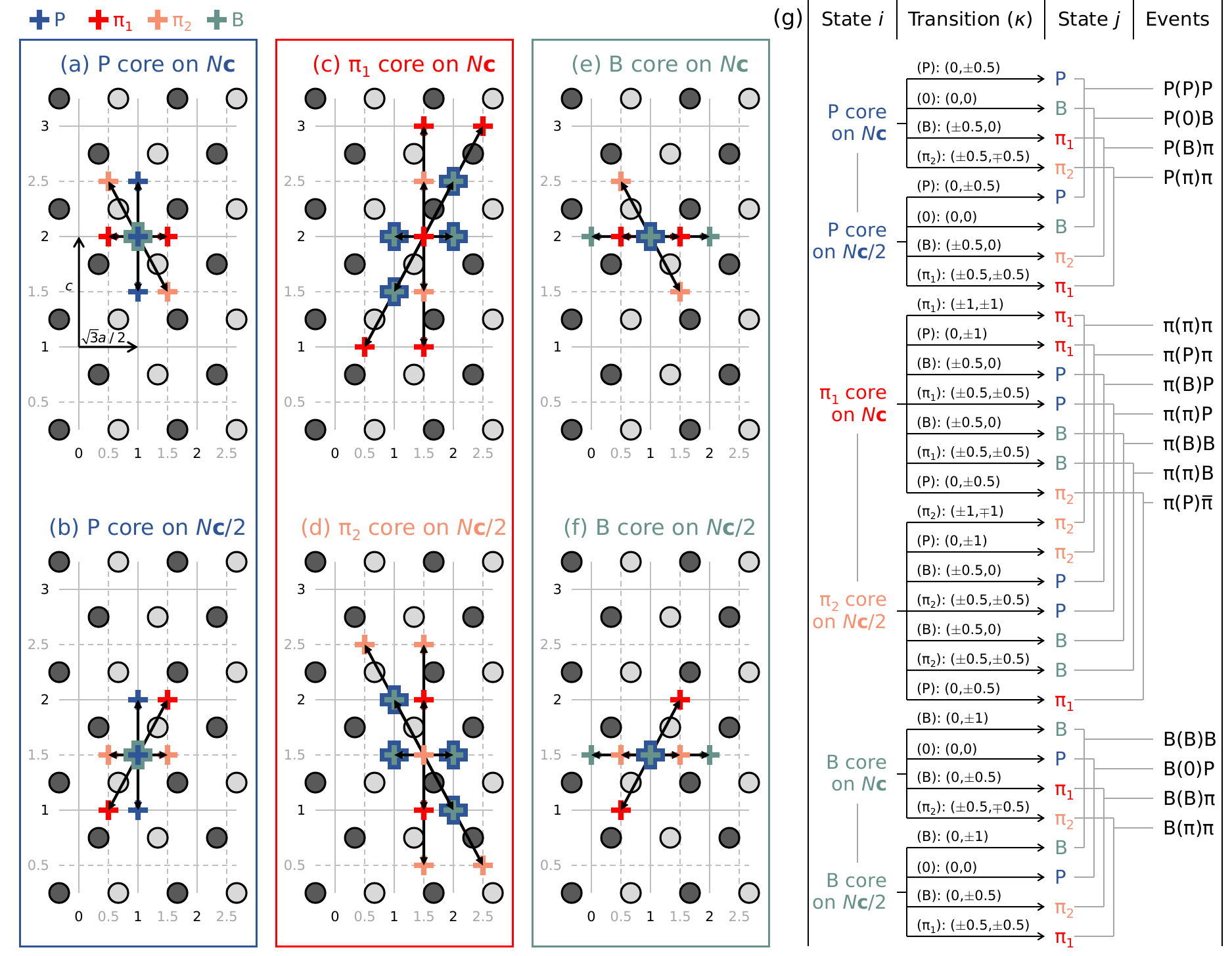}
\caption{\label{fig:events}
Kinetic events for  $\langle \mathbf{a}\rangle$ screw dislocation core dynamics.
(a)-(f) All possible events starting from a P core, a $\unslant{\pi}$ core or a B core. 
The dark gray and light gray circles denote  atoms on successive $(1\bar{2}10)$ planes. 
The vertical gray solid lines denote a series of $(\bar{2}020)$ planes; the vertical gray dashed lines are offset with respect to the vertical solid lines by $\sqrt{3}a/4$. 
The horizontal gray solid lines denote a series of $(0001)$ planes ($Nc$ planes); the horizontal gray dashed lines are offset with respect to the horizontal solid lines by $c/2$ (i.e., $Nc/2$ planes). 
The ``$+$'' symbols denote  dislocation core positions and their colors are consistent with those in the first column of (g). 
When two ``$+$'' symbols are located at the same site, we make one larger than the other for clarity.
(g) The 1$^\text{st}$ and 3$^\text{rd}$ columns show the starting and ending core structure for one event. 
The 2$^\text{nd}$nd column shows the core displacements corresponding to the transitions. 
Some events listed in the 1$^\text{st}$, 2$^\text{nd}$ and 3$^\text{rd}$ columns are connected by the gray lines indicating that they are equivalent in the sense that they have the same energy landscapes.  
The 4$^\text{th}$ column shows the irreducible events, where $i(\kappa)j$ denotes the transition from an $i$  to $j$ core by glide on plane $\kappa$. 
} 
\end{figure*} 

%%%%%%%%
\subsection{Kinetic Events}

%Based on the observations of the MD results, we conclude that there are two types of kinetic events during the motion of a dislocation core -- the transition of core structure and the glide of a core on a slip plane. 
The unit kinetic event during the motion of a dislocation core observed in the MD simulation can be abstracted as a transition. 
We label the states before and after a transition by $i$ and $j$, respectively, where $i,j \in \text{\{B, $\unslant{\pi}_1$, $\unslant{\pi}_2$, P\}}$. 
The $\unslant{\pi}_1$ and $\unslant{\pi}_2$ cores denote, respectively, the $\unslant{\pi}$ cores with the dissociation directions $\theta_{\unslant{\pi}} \approx 75^\circ$ and $105^\circ$ (they are symmetry-related). 
We use $\kappa$ to label the slip plane, i.e., $\kappa \in \text{\{B, $\mathrm{\pi}_1$, $\unslant{\pi}_2$, P, 0\}}$, where the $\unslant{\pi}_1$ and $\unslant{\pi}_2$ planes are, respectively, the $\unslant{\pi}'$ planes with the inclination angles $\phi \approx 61^\circ$ and $119^\circ$, and $\kappa =0$ denotes the transition which does not involve the change in core position. 
%We denote the transition of core structure from $i$ to $j$ as ``$i \to j$'' ($j\neq i$) and denote the glide of $i$ core on the $\alpha$ plane as ``$i \to i(\alpha)$''. 
%With the above notations, we can denote all events in a unified way by ``$i \to k$'', where $k\equiv j$ or $i(\alpha)$. 
We denote the transition from an $i$  to $j$ core involving glide on the $\kappa$ plane as ``$i(\kappa)j$''. 
An ``$i \ne j$'' event represents a transition of the core structure, while an ``$i = j$'' event represents glide of a core with no core structural transition. 
All  possible kinetic events observed in the MD simulations (Fig.~\ref{fig:set}) are shown schematically in Figs.~\ref{fig:events}a-f. 
These events, denoted  $i(\kappa)j$, are summarized in Fig.~\ref{fig:events}g. 
Some events listed in the 1st, 2nd and 3rd columns are equivalent. 
The 4th column shows the irreducible events. 
Note that $\bar{\unslant{\pi}}$ denotes the core symmetrically related to $\unslant{\pi}$; e.g., if $\unslant{\pi} = \unslant{\pi}_1$, then $\bar{\unslant{\pi}} = \unslant{\pi}_2$.  

%%%%%%%%
\subsection{Dislocation Core Free Energy}

\begin{figure}[t]
\includegraphics[width=0.88\linewidth]{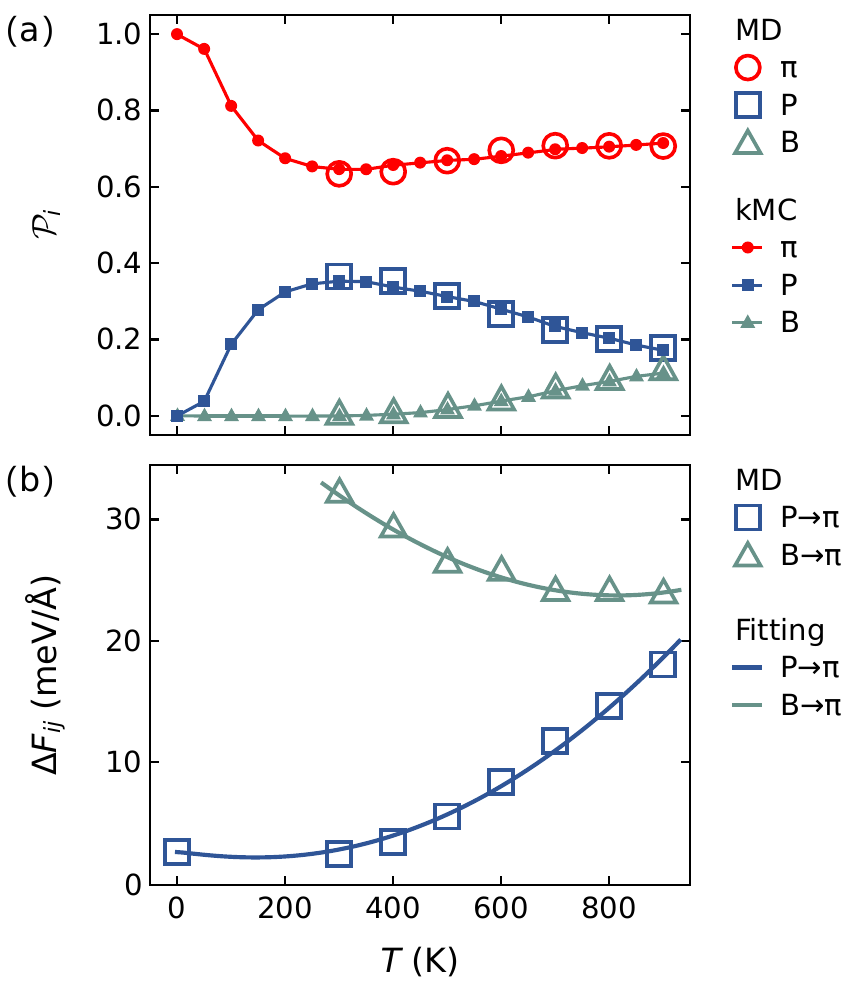}
\caption{\label{fig:probability}
The equilibrium probabilities of core structures and the free energy differences. 
(a) The probabilities of the $\unslant{\pi}$ (red circles), P (blue squares) and B (green triangles) cores obtained by the MD simulation (open symbols) and the kMC simulation (solid symbols). 
(b) The core energy differences, $\Delta F_{\txP\unslant{\pi}}$ (blue squares) and $\Delta F_{\txB\unslant{\pi}}$ (green triangles), obtained from the MD data in (a). 
The solid lines are the fitting results based on the formula Eq.~\eqref{DFijTABC}. 
}
\end{figure} 

From the MD results shown in Fig.~\ref{fig:set}c, we can extract the equilibrium probabilities $\{\mathcal{P}_i\}$ that a core is of type $i$ as a function of temperature. 
The open symbols in Fig.~\ref{fig:probability}a represent the MD data $\{\mathcal{P}_i\}$. 
$\unslant{\pi}$  is the most probable core structure at all temperatures,$\mathcal{P}_{\unslant{\pi}}(T)>0.6$. 
At $T < 300$~K, the B core is not observed in the MD and at $T> 300$~K, the B core occurs with very low probability. 

In thermal equilibrium at  temperature $T$, the probability of finding the $i$ core is well-described by a Boltzmann distribution:  
$\mathcal{P}_i \propto \exp\left(-F_i \ell \middle/ k_\txB T\right)$, where $F_i$ is the free energy of the $i$ core (energy per length), $\ell$ is the length of the dislocation line, and $k_\txB$ is the Boltzmann constant.  
The free energy difference between the $i$ and $j$ cores is related to their probability ratio:
\begin{equation}\label{prob_ratio}
\Delta F_{ij}
\equiv F_j - F_i
= \frac{k_\txB T}{\ell} 
\ln\left(\frac{\mathcal{P}_i}{\mathcal{P}_j}\right)
\text{ with $i\ne j$}. 
\end{equation}
Since the elastic energy is the same for all  core structures, $\Delta F_{ij}$ represents the core energy difference.
The core energy difference $\Delta F_{ij}$ at each temperature can be obtained from the MD data $\{\mathcal{P}_i\}$ (reported in Fig.~\ref{fig:probability}a) and Eq.~\eqref{prob_ratio}. 
Note that although the 10-20 step partial relaxations reduce the potential energy of the system, these energies are not of interest. Rather the partial relaxation simply aids the identification of the inherent core structures. The important energy differences $\Delta F_{ij}$ are determined  based on the relative probabilities of different dislocation core structures -- which are unaffected by the partial relaxations. 
The core energy differences, $\Delta F_{\txP\unslant{\pi}}$ and $\Delta F_{\txB\unslant{\pi}}$, are shown as open symbols in Fig.~\ref{fig:probability}b. 
We can fit the data $\Delta F_{ij}$ with an empirical relationship of the form:
\begin{equation}\label{DFijTABC}
\Delta F_{ij}(T)
= A + B T \ln T + C T^2, 
\end{equation}
where $A$, $B$ and $C$ are  parameters. The rationale for the form of this fitting relation is discussed in the SI.
For $\Delta F_{\txP\unslant{\pi}}$, $A=E_\txP - E_{\unslant{\pi}} = 2.6$~meV/{\AA}, where $E_\txP$ and $E_{\unslant{\pi}}$ are the energies of the P and $\unslant{\pi}$ cores at 0~K. 
The fitted curves are shown as solid lines in Fig.~\ref{fig:probability}b and the  parameters are  in Table~\ref{tab:DFij}. 

\begin{table}
\caption{\label{tab:DFij} Fit parameters for the free energy difference between  cores $i$ and $j$, $\Delta F_{ij}$, as per  Eq.~\eqref{DFijTABC}. }
\begin{ruledtabular}
\begin{tabular}{lll}
Parameters & $\Delta F_{\txP \unslant{\pi}}$ & $\Delta F_{\txB \unslant{\pi}}$ \\
\hline
$A$ (meV\,{\AA}$^{-1}$) &  $2.6$ &  $41.7 \pm 1.25$ \\
$B$ ($10^{-3}$~meV\,{\AA}$^{-1}$\,K$^{-1}$) &  $-1.65\pm 0.300$ &  $-7.65\pm 0.833$ \\
$C$ ($10^{-5}$~meV\,{\AA}$^{-1}$\,K$^{-2}$) &  $3.24 \pm 0.266$ &  $3.59 \pm 0.506$ \\
\end{tabular}
\end{ruledtabular}
\end{table}

%%%%%%%%
\subsection{Core Dynamics Parameterization}
\label{sect:coredynamics}

The basic kinetic parameters for dislocation core dynamics are the frequencies of all events as a function of  temperature. 
Unfortunately, it is impractical to deduce these frequencies directly from the MD results. 
Atomic vibrations are inevitable in MD simulations at finite temperatures; removing these by thermal averaging or quenching in order to unambiguously recognize each core event (defined in Fig.~\ref{fig:events}) requires  artificial criteria. 
%For example, we need to set a rule to judge if a dislocation core has finished a glide event or it is just in the process of thermal vibration. 
Sampling frequency also makes this impractical: if the sampling frequency is too high, the thermal vibration issue will be severe; if it is too low, we will miss some kinetic events. 
Here,  we sidestep the frequency issue, as explained  in this section. 
%On the other hand, the parameter space can be largely reduced when a physical model is assumed. 
%For example, if we assume that the temperature dependence of the core dynamics follows the Arrhenius law, we can choose the attempt frequencies and the intrinsic energy barriers, which are temperature independent, as the kinetic parameters. 
In short, we fit the MD data using harmonic transition state theory (HTST) with  temperature-independent parameters. 
The general parameterization steps are  
\begin{itemize}[font=\itshape]
\item[S1:] (Prediction) Guess a set of frequencies for all kinetic events $\{\nu_{i(\kappa)j}\}$ at each temperature (see Methods). % studied in the MD simulations. 

\item[S2:] (kMC) At each temperature, perform a kMC simulation (see Methods) with  frequencies $\{\nu_{i(\kappa)j}\}$ to compute the core mean squared displacement (MSD), the mean squared angular displacement (MSAD) of the core dissociation direction, and the probability of occurrence of each core  structure $\{\mathcal{P}_i\}$.

\item[S3:] (Optimization) Optimize $\{\nu_{i(\kappa)j}\}$ such that the MSD, MSAD and $\{\mathcal{P}_i\}$ obtained from the kMC are consistent with the MD results at all temperatures (see Methods). 

\item[S4:] (Reduction) From $\{\nu_{i(\kappa)j}\}$, extract the fundamental kinetic parameters (attempt frequencies and  intrinsic energy barriers) based on HTST.  
\end{itemize}

%%%
%\subsubsection*{S2: Kinetic Monte Carlo simulation}\label{KMC_simulation}

\begin{figure*}[t]
\includegraphics[width=0.95\linewidth]{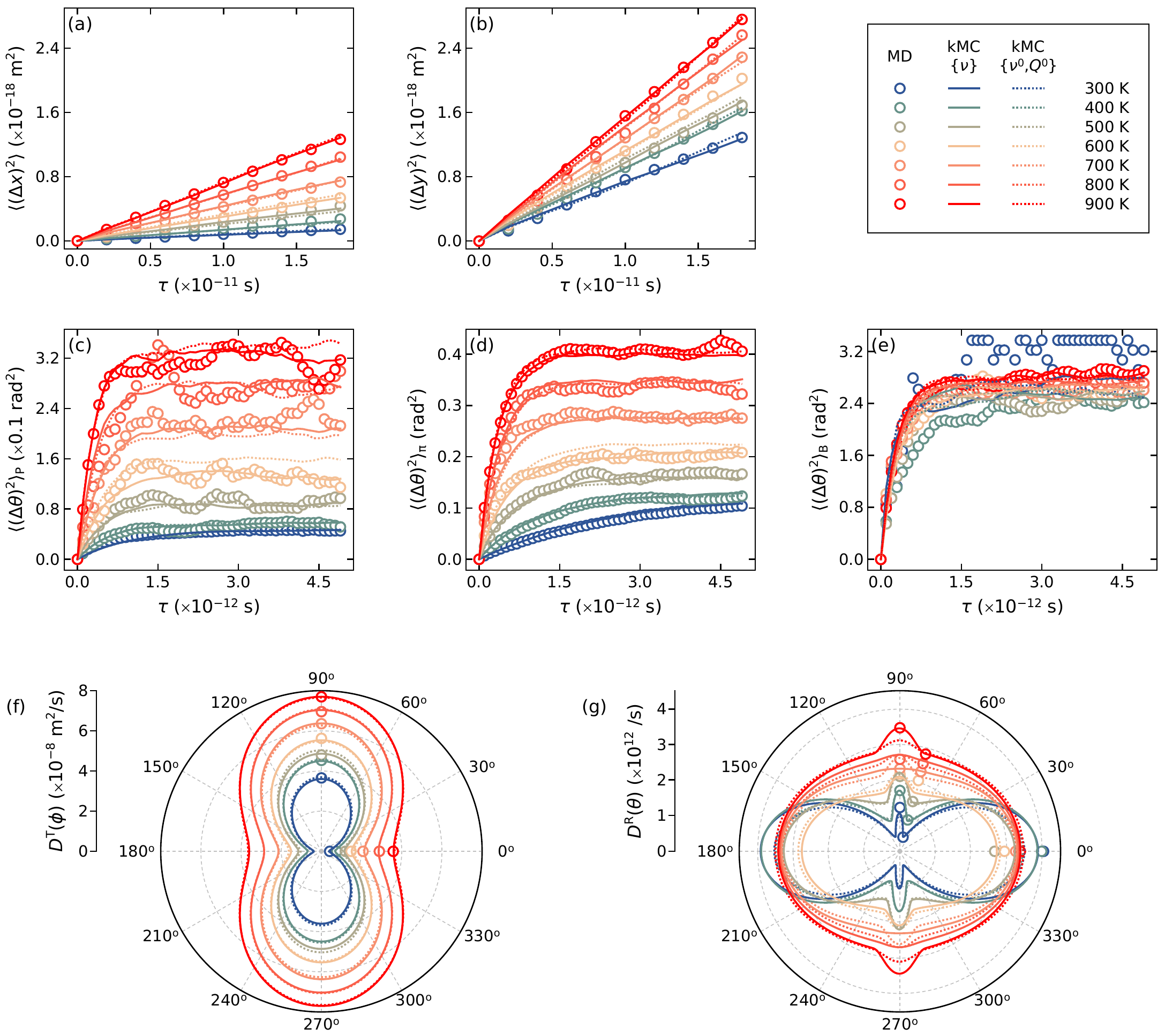}
\caption{\label{fig:msd}
Mean squared displacement and diffusion coefficients. 
(a) and (b) show the mean squared displacements (MSDs) of a dislocation core in the $x$- and $y$-directions at different temperatures. 
(c), (d) and (e) show the mean squared angular displacements (MSADs) of a core dissociation direction starting from $\theta_\txP$, $\theta_{\unslant{\pi}}$ and $\theta_\txB$. 
(f) Translational diffusion coefficient as a function of glide direction, $D^\txT(\phi)$ at different temperatures. 
(g) Rotational diffusion coefficient as a function of dissociation direction, $D^\txR(\theta)$ at different temperatures. 
In all figures, the circles denote the MD data, the solid lines denote the results of kMC simulations based on the optimized frequencies $\{\nu_{i(\kappa)j}\}$, and the dotted lines denote the results of kMC simulations based on the fitted parameters $\{\nu_{i(\kappa)j}^0\}$ and $\{Q_{i(\kappa)j}^0\}$. 
}
\end{figure*}

%%%
%\subsubsection*{S3: Optimization}

The mean squared displacement MSD and mean squared angular displacement MSAD (\textit{S2} and \textit{S3}) are defined as follows. 
At a particular temperature, the mean squared displacement MSD in the $\mathbf{e}_x$- and $\mathbf{e}_y$-directions are defined as
\begin{align}
&\langle (\Delta x)^2\rangle(\tau)
= \left\langle [x(t+\tau) - x(t)]^2\right\rangle, 
\nonumber\\
&\langle (\Delta y)^2\rangle(\tau)
= \left\langle [y(t+\tau) - y(t)]^2\right\rangle, 
\end{align}
where $\langle \cdot \rangle$ denotes the average over $t$. 
The circles in Figs.~\ref{fig:msd}a and b show the MSDs, $\langle (\Delta x)^2\rangle$ and $\langle (\Delta y)^2\rangle$, obtained from the MD simulations under different temperatures. 
We find that  core glide along the $\mathbf{e}_y$-axis (P plane) is, in general, faster than  glide along the $\mathbf{e}_x$-axis (B plane). 
At all temperatures, each MSD is approximately a linear function of $\tau$. 
The translational diffusion coefficients in the $\mathbf{e}_x$- and $\mathbf{e}_y$-directions are 
\begin{equation}
D_x^\txT = \frac{1}{2} \frac{\ud \langle(\Delta x)^2\rangle}{\ud \tau}
\quad\text{and}\quad
D_y^\txT = \frac{1}{2} \frac{\ud \langle(\Delta y)^2\rangle}{\ud \tau}. 
\end{equation}
One goal of optimizing the frequencies $\{\nu_{i(\kappa)j}\}$ is to ensure that the values of $D_x^\txT$ and $D_y^\txT$ from  kMC and MD simulations match. 
The mean squared angular displacement MSAD for  core dissociation $i$ is  
\begin{equation}\label{MSAD}
\langle(\Delta\theta)^2\rangle_i(\tau)
= \left\langle [\theta(t+\tau) - \theta(t)]^2 \right\rangle_i, 
\end{equation}
where the subscript ``$i$'' denotes that in the average $\theta(t=0) = \theta_i$, i.e., the dissociation direction of core $i$. 
The circles in Figs.~\ref{fig:msd}c, d and e show the MSADs, $\langle(\Delta\theta)^2\rangle_\txP$, $\langle(\Delta\theta)^2\rangle_{\unslant{\pi}}$ and $\langle(\Delta\theta)^2\rangle_\txB$, obtained from the MD simulations at different temperatures. 
The MSAD, at each temperature, is well fitted by the function: 
\begin{equation}
\langle(\Delta\theta)^2\rangle_i(\tau)
= 2\left(1 - e^{-D^\txR_i \tau}\right), 
\end{equation}
where $D^\txR_i$ ($i = \txP, \unslant{\pi}, \txB$) is the rotational diffusion coefficient about  dissociation angle $\theta_i$~\cite{jain2017diffusing}. 
In \textit{S3}, the frequencies $\{\nu_{i(\kappa)j}\}$ are trained for the best match between the core probabilities ($\{\mathcal{P}_i\}$), the translational diffusion coefficients ($D^\txT_x$ and $D^\txT_y$) and the rotational diffusion coefficients ($D^\txR_\txP$, $D^\txR_{\unslant{\pi}}$ and $D^\txR_\txB$) obtained by kMC and the MD results. 
The solid lines in Figs.~\ref{fig:msd}a-c show the MSDs and MSADs obtained from  kMC simulations with  optimized $\{\nu_{i(\kappa)j}\}$ for different temperatures. 
The kMC and MD results are in excellent agreement.
%%%
%\subsubsection*{S4: Parameter space reduction}

We now turn to the parameter space reduction in \textit{S4}.
The frequencies $\{\nu_{i(\kappa)j}\}$ are obtained via \textit{S1-3} for the MD simulation temperatures. 
In principle, dislocation core dynamics at other temperatures may also be obtained from MD simulation at such temperatures and repeating \textit{S1-3}. 
The computational resources required for these MD simulations and the optimization process limit the applicability of this  approach. 
We resolve this issue based upon a set of additional assumptions. 

\begin{figure}[t]
\includegraphics[width=0.95\linewidth]{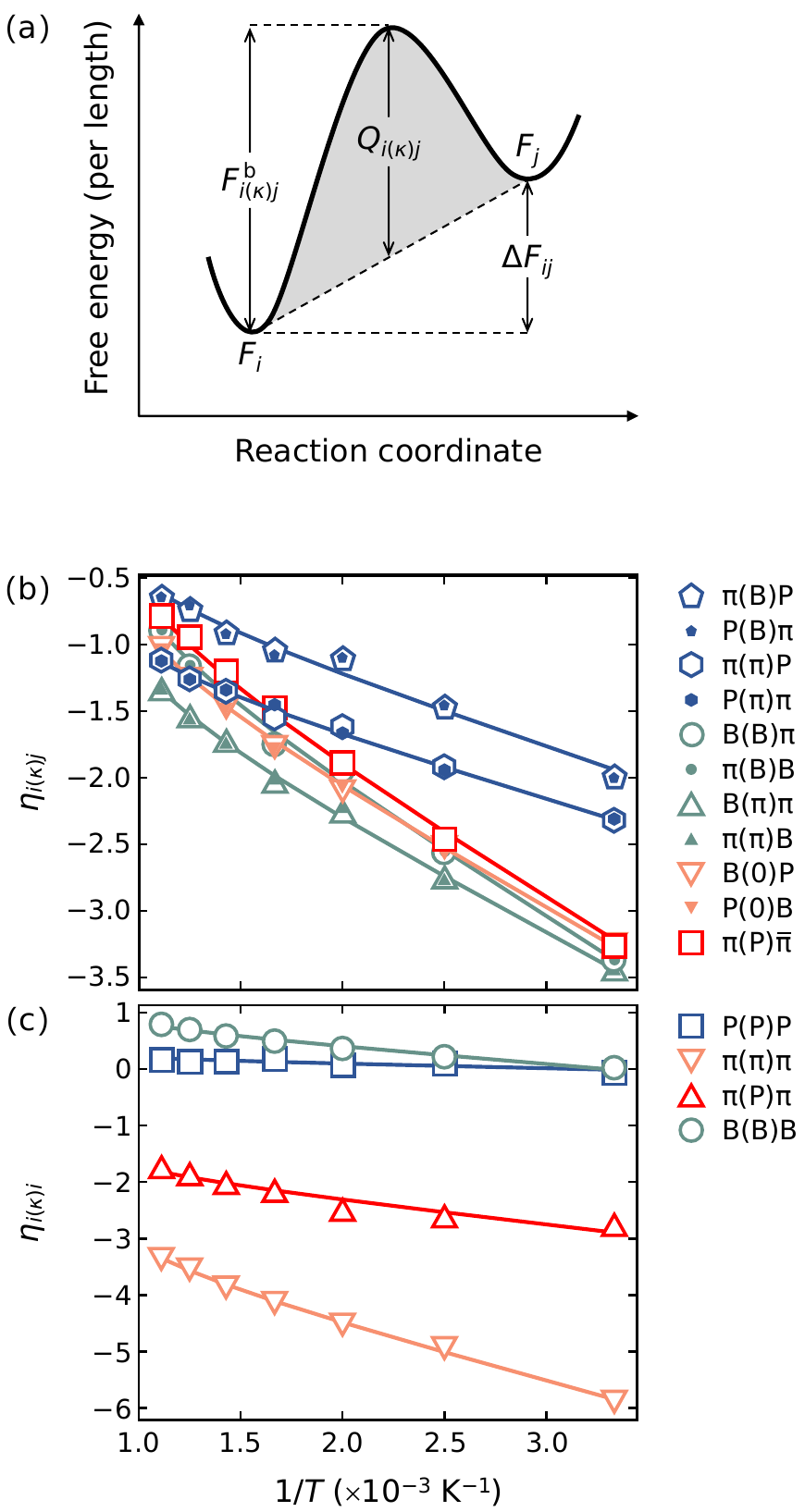}
\caption{\label{fig:fit}
Kinetic model and coefficients. 
(a) Schematic of the energy landscape for the $i(\kappa)j$ event. 
$F_i$ and $F_j$ are the free energies of the $i$ and $j$ cores; $\Delta F_{ij} = F_j - F_i$ and $\Delta F_{ij}=0$. 
$F_{i(\kappa)j}^\txb$ is the total free energy barrier and $Q_{i(\kappa)j}$ is the intrinsic barrier for $i(\kappa)j$. 
(b) $\eta_{i(\kappa)j}$ (Eq.~\eqref{etaik}) vs.  inverse  temperature. 
(c) $\eta_{i(\kappa)i}$ (Eq.~\eqref{etaik} for  glide events) vs. the inverse temperature. 
In (b) and (c), symbols and solid lines denote MD data and fitting results (Eqs.~\eqref{etaik} and \eqref{QikT}).  
}
\end{figure} 

Consider the schematic energy landscape for a kinetic event $i(\kappa)j$  in Fig.~\ref{fig:fit}a. 
Two local minima correspond to the $i$ and $j$ core free energies, $F_i$ and $F_j$. 
%When $k=j$, this event is the transition of core structure; when $k = i(\alpha)$, this event is the glide of $i$ core on the $\alpha$ plane. 
The core energy difference is $\Delta F_{ij}$ (Eq.~\eqref{prob_ratio});  for a core glide event, $\Delta F_{ii} = 0$. 
The total free energy barrier for  $i(\kappa)j$  is denoted  $F_{i(\kappa)j}^\txb$. 
The intrinsic free energy barrier for  $i(\kappa)j$  is $Q_{i(\kappa)j}$ and, in principle, $Q_{i(\kappa)j} = Q_{j(\kappa)i}$. 
%$Q_{ik}$ represents the intrinsic ``transition'' or ``glide'' barrier when $k=j$ or $k=i(\alpha)$. 
The total free energy barrier is commonly approximated as
\begin{equation}\label{eq:assumption_core_barrier}
F_{i(\kappa)j}^\txb 
= Q_{i(\kappa)j} + \Delta F_{ij} / 2,  
\end{equation}
where the factor $1/2$ is valid when $\Delta F_{ij} \ll Q_{i(\kappa)j}$~\cite{Cai1999kmcSi}. 
While other reasonable proposals for $F_{i(\kappa)j}^\txb$ are possible, our fitting results, below, suggest that Eq.~\eqref{eq:assumption_core_barrier} reproduces the MD results. 

The frequency of event $i(\kappa)j$ can be expressed as 
\begin{equation}\label{fiknuikexp}
\nu_{i(\kappa)j}
= \nu_{i(\kappa)j}^0 
\exp\left(-\frac{F_{i(\kappa)j}^\txb \ell}{k_\txB T}\right) 
\end{equation}
based on  HTST~\cite{eyring1935activated,vineyard1957frequency}, where $\nu_{i(\kappa)j}^0$ is an attempt frequency that includes the effect of barrier recrossing. 
%The frequencies $\{\nu_{i(\alpha)j}\}$ can be measured from the MD results at each temperature. 
Equation~\eqref{fiknuikexp} can be rewritten as% (with Eqs.~\eqref{prob_ratio} and \eqref{eq:assumption_core_barrier}) 
\begin{align}\label{etaik}
\eta_{i(\kappa)j}(T^{-1})
&\equiv \ln \left(
\nu_{i(\kappa)j} \sqrt{\mathcal{P}_i / \mathcal{P}_j}
\right)
\nonumber\\
&= 
\ln \nu_{i(\kappa)j}^0
- \frac{Q_{i(\kappa)j} \ell}{k_\txB T}.  
\end{align}
This shows that $\eta_{i(\kappa)j}$ is a function of the inverse temperature $T^{-1}$.
If $\{\nu_{i(\kappa)j}^0\}$ and $\{Q_{i(\kappa)j}\}$ are constant with respect to temperature, they can be obtained by linear fitting. 
However, we find that the fitting quality is  improved by allowing the intrinsic free energy barrier to be temperature-dependent and of the form:
\begin{equation}\label{QikT}
Q_{i(\kappa)j}
= Q_{i(\kappa)j}^0 \left[
1 - \left(\frac{T}{T_0}\right)^q
\right],
\end{equation}
where $Q_{i(\kappa)j}^0$ is the intrinsic barrier at 0~K, $T_0 = 1250$~K is the HCP-BCC transition temperature for this DP potential~\cite{wen_Ti}, and we assign $q = 3$ to give best fit to all the data.  
Equation~\eqref{QikT} is proposed based on  the observations that (i) the free energy barrier  decrease with increasing temperature and (ii) when the HCP phase becomes unstable/metastable, the transition between the core structures is barrierless. 
In this way, the kinetic parameters can be fitted to the MD data; i.e., $\{\nu_{i(\kappa)j}^0\}$ and $\{Q_{i(\kappa)j}^0\}$. 

Figures~\ref{fig:fit}b and c show the fitting results for Eqs.~\eqref{etaik} and \eqref{QikT} (symbols/lines are the MD data/fits). 
As expected, the  $i(\kappa)j$ and $j(\kappa)i$ data coincide. 
Among the transition events (Fig.~\ref{fig:fit}b), the transitions between P and $\unslant{\pi}$ cores are the most frequent and associated with the lowest energy barriers. 
As expected, direct transitions between P and B cores are rare. 
Among the glide events (Fig.~\ref{fig:fit}c), the glide on B plane is associated with the lowest barrier. 
But glide on B plane is rare since the probability of the B core is very low (Fig.~\ref{fig:probability}a). 
Pyramidal glide (via $\unslant{\pi}(\unslant{\pi})\unslant{\pi}$) is much less frequent and associated with a much higher barrier than  prismatic glide (via P(P)P). 
This means that pyramidal glide is rarer than  prismatic glide; in qualitative agreement with the DFT 0~K glide barriers~\cite{Clouet2015locking} (DFT predicts the Peierls barrier for P(P)P glide and $\pi(\pi)\pi$ glide as 11.4 meV\,\AA$^{-1}$ and 0.4 meV\,\AA$^{-1}$, while the values from our simulations are 14.1 meV\,\AA$^{-1}$ and 1.14 meV\,\AA$^{-1}$).

\begin{table}
\caption{\label{tab:fitting}The kinetic parameters determined by fitting to the MD data for the core transition events and glide events. 
$\nu_{ik}^0$ is the attempt frequency in Eq.~\eqref{etaik}. 
$Q_{ik}^0$ is the intrinsic 0~K energy barrier in Eq.~\eqref{QikT}. 
}
\begin{ruledtabular}
\begin{tabular}{lll}
Events $i(\kappa)j$ & $Q_{i(\kappa)j}^0$~(meV\,{\AA}$^{-1}$) & $\nu_{i(\kappa)j}^0$~(ps$^{-1}$) \\
\hline
\multicolumn{3}{l}{$i\ne j$: core transition} \\
B(B)$\unslant{\pi}$ &
$13.9 \pm 0.444$ & $0.782 \pm 0.0450$ \\
B(0)P &
$12.6 \pm 0.172$ & $0.636 \pm 0.0141$ \\
$\unslant{\pi}$(B)P &
$7.47 \pm 0.447$ & $0.769 \pm 0.0445$ \\
$\unslant{\pi}$(P)$\bar{\unslant{\pi}}$ &
$13.7 \pm 0.209$ & $0.871 \pm 0.0236$ \\
$\unslant{\pi}$($\unslant{\pi}$)P &
$6.69 \pm 0.125$ & $0.442 \pm 0.00718$ \\
B($\unslant{\pi}$)$\unslant{\pi}$ &
$11.8 \pm 0.221$ & $0.457 \pm 0.0131$ \\
\hline
\multicolumn{3}{l}{$i=j$: core glide} \\
B(B)B &
$4.33 \pm 0.237$ & $2.61 \pm 0.0800$ \\
$\unslant{\pi}$($\unslant{\pi}$)$\unslant{\pi}$ &
$14.1 \pm 0.315$ & $0.0689 \pm 0.00281$ \\
$\unslant{\pi}$(P)$\unslant{\pi}$ &
$6.06 \pm 0.760$ & $0.216 \pm 0.0213$ \\
P(P)P &
$1.14 \pm 0.295$ & $1.18 \pm 0.0488$ \\
\end{tabular}
\end{ruledtabular}
\end{table}

%%%%%%%%%%%%%%%%%%%%%%%%%%%%%%%%%%%%%%%%%%%%%%%%%%%%%%%%%%
\section{Kinetic Monte Carlo simulation results}

The model and parameterization of the core dynamics described are the key for understanding and predicting dislocation dynamics. 
Here, we apply this model to  $\langle \mathbf{a} \rangle$ screw dislocation dynamics in Ti via kinetic Monte Carlo (kMC) simulations (see Methods). 
The only modification is that the kinetic event frequencies  $\{\nu_{i(\kappa)j}\}$ are determined through  Eqs.~\eqref{eq:assumption_core_barrier}, \eqref{fiknuikexp} and \eqref{QikT}. 

%%%%%%
\subsection{Random Walk of a Dislocation Core}

The equilibrium  core structure probabilities $\{\mathcal{P}_i\}$ were computed via kMC and compared with those obtained from MD, as shown in Fig.~\ref{fig:probability}a. 
At temperatures $\ge 300$~K, the kMC results show excellent agreement with MD. 
$\mathcal{P}_{\unslant{\pi}}$ and $\mathcal{P}_\mathrm{P}$ exhibit a minimum and maximum near 300~K.
Beyond this temperature both $\mathcal{P}_{\unslant{\pi}}$ and $\mathcal{P}_\mathrm{B}$ increase, while $\mathcal{P}_\mathrm{P}$ decreases with  temperature $T$. 
%Starting from around 0.35 and 0 at 300~K, the values of $
$\mathcal{P}_\mathrm{P}$ and $\mathcal{P}_\mathrm{B}$ are approximately equal  near 0 and 900~K.
%close when heated and finally reach around 0.19 and 0.11 at 900~K. 
%Meanwhile, $\mathcal{P}_{\unslant{\pi}}$ increases slightly from around 0.62 to 0.71. 
No MD is available below  300~K where it is difficult to obtain valid statistics; here, we only show  kMC data from Eq.~(\ref{DFijTABC}). 
At 0~K, based on the fact that $\unslant{\pi}$ core is energetically favorable, $\mathcal{P}_{\unslant{\pi}}$ should be 1 and $\mathcal{P}_\mathrm{P}$ and $\mathcal{P}_\mathrm{B}$ should be 0. 
$\mathcal{P}_\mathrm{B}\approx0$ for $T\leq 300$~K, which is consistent with the MD observation that the  B core is unstable. 

The MSDs and MSADs  computed via MD and  kMC simulations are compared in Fig.~\ref{fig:msd}a-e. 
In general, the kMC simulations  reproduce the MD results well, except for the MSADs corresponding to core dissociation  starting from $\theta_\txB$ at low temperatures (Fig.~\ref{fig:msd}e) (this is associated with limited sampling of the B core in the MD simulations). 
Due to larger accessible timescale in kMC compared with MD, the kMC simulations provide  smooth curves at low computational cost  (compared with MD). 
The glide-direction dependent translational diffusion coefficient can be constructed as  
\begin{equation}
D^\mathrm{T}(\phi) 
= D^\mathrm{T}_x \cos^2\phi + D^\mathrm{T}_y \sin^2\phi;
\end{equation}
see Fig.~\ref{fig:msd}f. 
$D^\mathrm{T}(\phi)$ is elongated in the $\mathbf{e}_y$-direction, indicating that a dislocation core moves fast along the P plane and slowly along the B plane; this is consistent with the MD trajectories in Fig.~\ref{fig:set}a. 
The dissociation-angle-dependent rotational diffusion coefficient can be constructed as 
\begin{equation}
D^\mathrm{R}(\theta) 
= D^\mathrm{R}_i\cos^2\left(\frac{\pi}{2}\frac{\theta-\theta_i}{\theta_j-\theta_i}\right)
+ D^\mathrm{R}_j\sin^2\left(\frac{\pi}{2}\frac{\theta-\theta_i}{\theta_j-\theta_i}\right), 
\end{equation}
where $(i,j)=(\txB, \unslant{\pi})$ or $(\unslant{\pi}, \txP)$; see Fig.~\ref{fig:msd}g. 
$D^\txR(\theta)$ measures the rotation rate for cores initially oriented  at angle $\theta$. 
Figure~\ref{fig:msd}g shows that $D^\txR(\theta)$ is highly anisotropic at low temperatures and becomes more isotropic as temperature increases. 
At all temperatures (except for 900~K), $D^\txR$ is  maximum at $\theta = 0$ which corresponds to the B core. 
This is consistent with the fact that B core is not energetically favorable and tends to transform into other cores. 
$D^\txR$ is a minimum at  ${\sim}\theta = 75^\circ$;  corresponding to the $\pi$ core. 
This indicates that the most stable core is $\pi$.

%%%%%%
\subsection{Stress-Driven Dislocation Motion}

\begin{figure*}[t]
\includegraphics[width=0.7\linewidth]{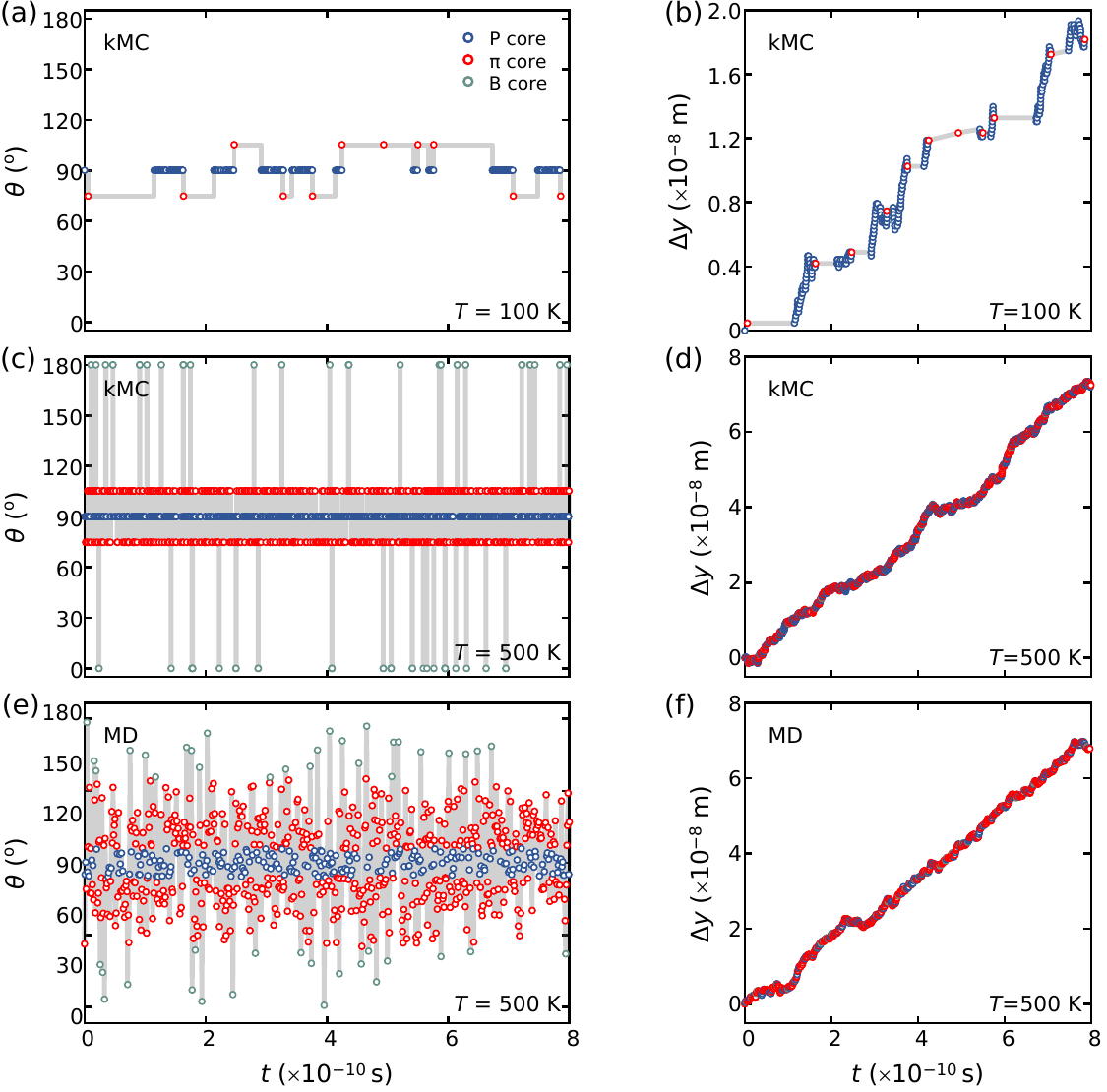}
\caption{\label{fig:distribution_mobility}
Temporal evolutions of the dislocation core dissociation angle ($\theta$) and the core displacement along P plane ($\Delta y$). 
(a) and (b) The core trajectories under $\sigma_{yz}=$10~MPa at 100~K predicted by kMC. 
(c) and (d) The core trajectories under $\sigma_{yz}=$60~MPa at 500~K predicted by kMC. 
(e) and (f) The core trajectories under $\sigma_{yz}=$60~MPa at 500~K obtained by MD. 
In all figures, blue, red and green circles label the P, $\unslant{\pi}$ and B cores, respectively.  
}
\end{figure*}
%%%
%\subsubsection{Stress dependent kinetic model}

The kMC model is sufficiently flexible to simulate dislocation dynamics under an externally applied stress.  
We apply the  approach developed by Ivanov and Mishin~\cite{mishin2008prb}. %(they focused on stress-driven grain boundary migration. 

The major assumption is that the applied stress  influences the dislocation glide barrier through the resolved shear stress (RSS), but not the energy barrier for the core transition itself. 
Hence, this model does not fully capture the non-Schmid effect~\cite{christian1983some,vsestak1965asymmetry,duesbery1998plastic,ito2001atomistic} (see below).

We assume that the free energy landscape for $i$ core glide   on the $\kappa$ plane has the form of $F_{i(\kappa)i}(r) = (Q_{i(\kappa)i}/2)[1 - \cos(2\pi r/L_{i(\kappa)i})]$, where $r$ is a slip distance, $Q_{i(\kappa)i}$ is the glide barrier (Eq.~\eqref{QikT}) and $L_{i(\kappa)i}$ is a lattice period in the slip direction on the $\kappa$ plane. 
Applying an external stress $\boldsymbol{\sigma}$ creates Gibbs free energy landscape  $G_{i(\kappa)i}(r) = F_{i(\kappa)i}(r) - f_\kappa r$, where $f_\kappa$ is the Peach-Koehler (PK) force: % resolved onto the slip direction $\mathbf{s}_\kappa$ on the $\kappa$ plane: 
%\begin{equation}
$f_\kappa = 
(\boldsymbol{\sigma} \mathbf{b}) \times \boldsymbol{\xi} \cdot \mathbf{s}_\kappa
= \tau_\kappa b$.
%\end{equation}
With our sample geometry, Fig.~\ref{fig:model}a, $\boldsymbol{\xi} = \mathbf{e}_z$ is the line direction, $\mathbf{b} = b\mathbf{e}_z$ is the Burgers vector, $\mathbf{s}_\kappa = \cos\phi_\kappa\mathbf{e}_x + \sin\phi_\kappa\mathbf{e}_y$ is the slip direction on the $\kappa$ plane with  inclination angle $\phi_\kappa$, and $\tau_\kappa \equiv -\sigma_{xz}\sin\phi_\kappa + \sigma_{yz}\cos\phi_\kappa$ is the RSS on the $\kappa$ plane. 
Then, the Gibbs free energy barrier for the forward/backward glide is 
\begin{align}\label{eq:stress_reduce_barrier}
G_{i(\kappa)i}^{\txb \pm}
&=
Q_{i(\kappa)i} \Bigg[
\sqrt{1-\left(\frac{\tau_\kappa}{\tau_{i(\kappa)i}^0}\right)^2}
\nonumber\\
&\mp 
\frac{\pi \tau_\kappa}{2 \tau_{i(\kappa)i}^0}
\left(1 \mp \frac{2}{\pi}\arcsin \frac{\tau_\kappa}{\tau_{i(\kappa)i}^0}\right) 
\Bigg], 
\end{align}
where $\tau_{i(\kappa)i}^0 \equiv \pi Q_{i(\kappa)i}/(L_{i(\kappa)i} b)$. 
The frequency of forward/backward glide event $i(\kappa)i$, $\nu_{i(\kappa)i}^{\pm}$, is  obtained from Eq.~\eqref{fiknuikexp} with  $F_{i(\kappa)i}^\txb$ replaced by $G_{i(\kappa)i}^{\txb\pm}$;  $\nu_{i(\kappa)i} = \nu_{i(\kappa)i}^+ - \nu_{i(\kappa)i}^-$. 
The detailed explanation of Eq.~\eqref{eq:stress_reduce_barrier} can be found in Ref.~\cite{mishin2008prb}. 
With this frequency as input, we perform kMC simulations of dislocation  motion under different stresses $\boldsymbol{\sigma}$ (see Methods).

%%%%%%%%%%%%%%%%%%%%%%%%%%%%%%%%%%
%\subsubsection{Locking-unlocking motion}

The kMC simulations show a locking-unlocking type of dislocation motion at low temperatures. 
Figures~\ref{fig:distribution_mobility}a and b show the temporal evolution of the dislocation core dissociation angle, $\theta$, and the core displacement in the $\mathbf{e}_y$-direction (P plane), $\Delta y$, at $T = 100$~K  under  shear stress $\sigma_{yz}=10$~MPa. 
When the dislocation core has a P core (blue circles), $\Delta y$ increases quickly; i.e., P-glide is fast. 
On  transformation to a $\unslant{\pi}$ core (red circles), the dislocation pauses while ``waiting'' to transform back to a P core (Fig.~\ref{fig:distribution_mobility}a) upon which glide restarts on the P plane (Fig.~\ref{fig:distribution_mobility}b). 
Hence, the  $\unslant{\pi}$ core is ``locked'' (does not glide) and the P core is ``unlocked'' (glides easily). 

The ``locking'' period predicted by kMC ($\sim 10^{-10}$~s) is much smaller than that observed in \textit{in situ} TEM straining experiments ($\sim8$~s)~\cite{Clouet2015locking}. 
There are two possible sources for this discrepancy. 
The TEM specimen is a thin foil, the free surfaces of which could provide strong drag on the dislocation. 
We have performed additional MD simulations, involving a dislocation line threaded at two free surfaces, to examine the drag effect  (see SI for the simulation settings and results). 
The simulation results confirm that the free surfaces significantly reduce the dislocation mobility by imposing severe restrictions on core transitions.
Second, our kMC simulations only study the motion of a short dislocation segment, rather than a long dislocation line.  
The differences in dislocation mobility between our work and TEM observations are attributed to the differences in dislocation line length. The motion of  long dislocation lines involves collective motion of many dislocation segments, thus a higher free energy barrier should be overcome during the glide process. 
Moreover, long dislocations may migrate via kink pair nucleation and propagation, which necessitate inclusion of  kink formation and  migration energy barriers in estimation of dislocation mobility. 
Multiple core structures will likely be found on a long  dislocation line. The interaction between these different core structures may also lower the dislocation mobility.
This, coupled with the free surface restraint on core transitions (see SI) explains why short dislocation segments are more mobile than long dislocation lines in TEM observations. 

To validate our kMC results, we simulated dislocation core motion at high temperature (500~K) at a high shear stress ($\sigma_{yz} = 60$~MPa) by both kMC and DP-based MD (note that the MD time scale only allows the study of fast dynamics which can be achieved at high temperatures and high driving forces). 
%A comparison was made between the kMC and MD results.  
Figures~\ref{fig:distribution_mobility}c and d show the evolution of $\theta$ and $\Delta y$ obtained through kMC simulations, while Fig.~\ref{fig:distribution_mobility}e and f show the same quantities under the same conditions from MD. 
The kMC and MD results are  consistent. 
At this high temperature, the locking-unlocking mechanism is not easily seen, although it effectively lowers the core velocity along the P plane. 

%%%
\subsection{Dislocation Mobility}

A shear stress $\boldsymbol{\sigma} = \sigma_{xz}(\mathbf{e}_x \otimes \mathbf{e}_z + \mathbf{e}_z \otimes \mathbf{e}_x)$ creates a PK force on the screw dislocation $\mathbf{f}=\sigma_{xz}b \mathbf{e}_x$; the dislocation  moves, on average, in the $\mathbf{e}_x$-direction, i.e., on the B plane. 
We found that the dislocation velocity on the B plane, $v_\txB$, is a linear  function of $\sigma_{xz}$ (see the kMC and MD data in SI). 
The  dislocation mobility on the B plane is $M_\txB = v_\txB/(\sigma_{xz}b)$. 
Alternatively, we may drive the dislocation motion on the P plane via shear stress $\boldsymbol{\sigma} = \sigma_{yz}(\mathbf{e}_y \otimes \mathbf{e}_z + \mathbf{e}_z \otimes \mathbf{e}_y)$ to obtain $M_\txP = v_\txP/(\sigma_{yz}b)$. 
If the energy barrier for a  core transition or glide event is large, the intrinsic free energy barrier, $Q_{i(\kappa)j}$, is the Peierls barrier and the dislocation motion is thermally activated. 
On the other hand, if it is small, $Q_{i(\kappa)j}$ cannot be interpreted as Peierls barrier; it is simply a parameter in the model which reproduces the frequencies obtained from MD. 
Dislocation mobility in the case of small energy barrier is phonon damping-controlled such that the viscous drag  coefficient is $M_\txB^{-1}$ or $M_\txP^{-1}$~\cite{HirthLothe}. 
Phonon damping is not explicitly modeled in kMC; rather it is captured through the parameterization of the frequency obtained from MD  (similar to the method reported in Ref.~\cite{Stukowski2015kmc}). 

\begin{figure}[t]
\includegraphics[width=0.99\linewidth]{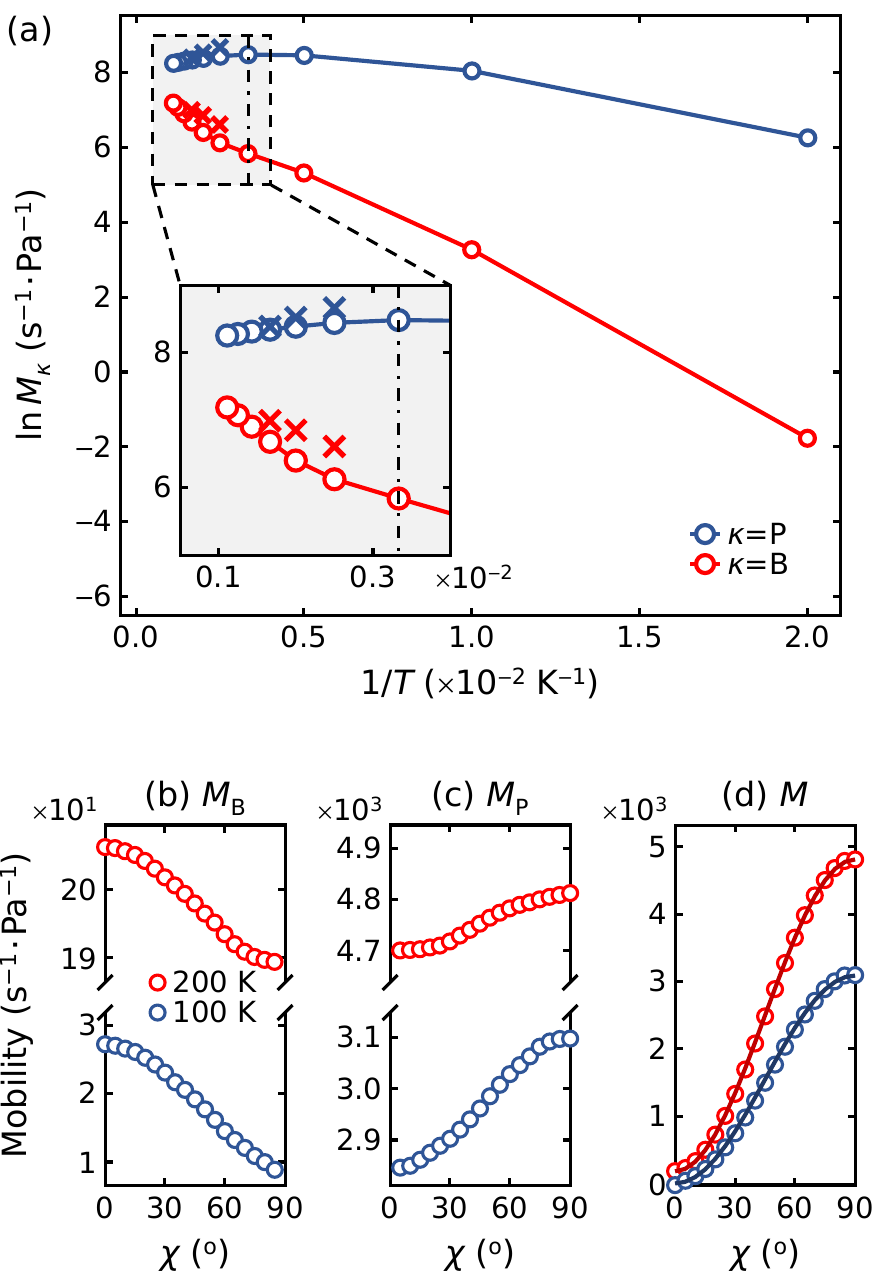}
\caption{\label{fig:mobility}
(a) Arrhenius plot of dislocation mobilities on the P plane  for $\sigma_{yz}$ (blue) and B plane for $\sigma_{xz}$ (red).
The circles and crosses are, respectively, the kMC and MD data. The dash-dotted line indicates $T=300$~K.
(c) and (d) The mobilities on the B and P planes ($M_\txB$ and $M_\txP$) as a function of  PK force direction $\chi$. 
(d) The scalar dislocation mobility as a function of $\chi$.
The symbols are kMC data while the solid lines are fits to Eq.~\eqref{Mpsi}. 
In (b)-(d) the blue and red circles correspond to 100~K and 200~K. 
}
\end{figure}

The temperature-dependencies of $M_\mathrm{B}$ and $M_\mathrm{P}$ are shown in Fig.~\ref{fig:mobility}. 
$M_\txB$ is much lower than $M_\txP$ at all temperatures, as suggested by Fig.~\ref{fig:msd}f (where the dislocation core diffusion coefficient is a minimum/maximum at $\phi = 0/90^\circ$. 
As shown in Fig.~\ref{fig:mobility}a and its inset, $M_\txB$  increases with temperature monotonically while $M_\txP$ exhibits a maximum at $T = 300$~K (indicated by the dash-dotted line). 
The decrease of $M_\txP$ and increase of $M_\txB$ above 300~K is consistent with the MD results, i.e., the crosses in Fig.~\ref{fig:mobility}a inset. 
MD simulations from the literature~\cite{Chu2020MDmob,Kuksin2008dp,Queyreau2011MDmob, Olmsted2005MDmob} always show that the dislocation mobility decreases (or equivalently, the viscous drag coefficient increases) with increasing temperature. 
However, we note that  MD results at low temperatures do not exist, since MD  timescales do not suffice. 
The decrease in mobility (increase in viscous drag coefficient) at high temperature is usually interpreted as a phonon drag/damping effect. 
However, our kMC results suggest that  glide on the P plane is also effectively damped by dislocation core transitions to B core. 
With increasing temperature, the B core is increasingly stable (Fig.~\ref{fig:probability}a) such that the  transition rate from the $\unslant{\pi}$ core to the B core increases (Fig.~\ref{fig:fit}b), leading to B-glide which contributes to zero motion on the  P plane.

We also investigated the effect of shear stress orientation and  dislocation mobility anisotropy. 
A PK force may be applied in different directions, by choice of the relative magnitudes of $\sigma_{xz}$ and $\sigma_{yz}$; i.e., $\mathbf{f} = \sigma_{xz}b\mathbf{e}_x + \sigma_{yz}b\mathbf{e}_y$.
The orientation angle (maximum resolved shear stress plane, MRSSP) is 
%\begin{equation}
$\chi 
= \arctan(f_\txP/f_\txB)
= \arctan(\sigma_{yz}/\sigma_{xz})$,
%\end{equation}
where $f_\mathrm{P}$ and $f_\mathrm{B}$ are the PK forces resolved on to the P and B planes.  
We calculated $M_\txP$ and $M_\txB$ for various values of $\chi$ under 100~K and 200~K; the results are shown in Figs.~\ref{fig:distribution_mobility}g and h. 
There is no surprise that $M_\mathrm{B}$ decreases and $M_\txP$ increases as $\chi$ increases from 0 to 90$^\circ$, and the high-temperature mobilities are higher than the low-temperature counterparts. 
The dislocation mobility can be generalized to a tensor, $\mathbf{M}$, defined by the relationship: $\mathbf{v} = \mathbf{M}\mathbf{f}$. 
The dislocation glide velocity component parallel to the PK force is $v = \hat{\mathbf{f}}\cdot\mathbf{v} = (\hat{\mathbf{f}}\cdot\mathbf{M}\hat{\mathbf{f}})f = Mf$, where $\hat{\mathbf{f}} = (\cos\chi, \sin\chi)^T$ is the direction of PK force and the scalar dislocation mobility is 
\begin{align}\label{Mpsi}
&M(\chi) 
\equiv \hat{\mathbf{f}}\cdot\mathbf{M}\hat{\mathbf{f}} 
\nonumber\\
&= M_{11}\cos^2\chi + M_{22}\sin^2\chi + 2M_{12}\sin\chi\cos\chi,
\end{align}
where $M_\txP = M(0) = M_{11}$ and $M_\txB = M(90^\circ) = M_{22}$. 
%We applied the PK force in various directions ($\chi$), 
We measured the velocity component parallel to the PK force ($v$) and calculated the scalar mobility ($M=v/f$) as a function of $\chi$; see Fig.~\ref{fig:mobility}d.  
Next, we extracted $M_{ij}$ by fitting $M$ vs. $\chi$ to Eq.~\eqref{Mpsi}; see the solid lines in Fig.~\ref{fig:mobility}d.
We find that $M_{11} = M_\txB(\chi=0)$, $M_{22} = M_\txP(\chi=90^\circ)$ and $M_{12}$ is negligibly small in comparison with $M_{11}$ and $M_{22}$. 
Since $M_{11} \ne M_{22}$, the dislocation velocity $\mathbf{v}$ is, in general, not in the same direction as the PK force $\mathbf{f}$. 
The off-diagonal component $M_{12}$  relates to how the PK force, resolved on B/P plane, influences dislocation glide on the P/B plane -- this is a non-Schmid effect~\cite{christian1983some}. 
$M_{12} \approx 0$ indicates that dislocation glide is well-described by the Schmid law in our kMC model.

%%%%%%%
\subsection{Limitations}\label{limitation}

\subsubsection{Non-Schmid effects}\label{sec:nonschmid}

The glide probabilities are deduced from dislocation random walk (i.e., with no driving force), from which we extracted the intrinsic glide barriers. 
Applied stress does not change the intrinsic glide barriers, but rather  biases the glide barriers according to Eq.~\eqref{eq:stress_reduce_barrier}.  
The Schmid effect is naturally included in this model, since the glide barrier will be lowered the most along the plane where the resolved shear stress is the maximum. 

We do not investigate non-Schmid effects in this study as this is not an intrinsic property. 
To capture  non-Schmid effects generally necessitates determination of how  the entire 6-dimensional stress tensor  alters the glide barrier.
More specifically, it is possible to repeat our sampling and analysis presented in this work as a function of  stress normal to the P-plane, $\pi$-plane or B-plane.  This is beyond the scope of this study.

\subsubsection{Long dislocation lines} 

The present simulations focused mainly on understanding the effect of intrinsic dislocation core properties/behavior on dislocation motion. 
Hence, all simulations were performed to short dislocation segments. 
While this helps obtain fundamental, intrinsic core behavior, it does not account for all aspects of dislocation dynamics.
Nevertheless, it is of practical interest to understand how a long, screw dislocation line moves in $\alpha$-Ti. 
The mechanism of the motion of a long dislocation line is associated with nucleation and propagation of kinks. 
Both kink nucleation and propagation necessarily involve the advance of local short dislocation segments. 
In this sense, the core properties extracted in this paper serve as essential input to such higher-level, long dislocation line models/simulations. 
 
A serious treatment of long dislocations should be multiscale. 
Even the extant long ($32b$) dislocation MD simulations (e.g., see Ref.~\cite{poschmann2022molecular}) have not resolved the size effect issue.  
At a high temperature ($k_\text{B}T$ exceeds the kink energy), a dislocation line will likely undergo thermal roughening (i.e., the fluctuation amplitude of a dislocation line scales with the size of system)~\cite{Aleinikava_2010}. 
If so, the size effect cannot be overcome by any finite length scale MD simulation. 
A possible strategy is to incorporate the intrinsic dislocation core properties as inputs for a multiscale method, such as kMC, rather than to simulate a long dislocation directly by MD.
To do this, additional information is required (e.g., double kink formation  and  migration barriers or migration velocities); these may be obtained either directly from MD simulations or by derivation. 
For example, Edagawa's line tension model~\cite{Edagawa1997kink} provides a reasonable description of variations of the double-kink nucleation energy, which can also be directly obtained from atomistic simulation by modeling a kink structure~\cite{xu1998kink}. 
The kink migration barrier may be obtained from atomistic simulation for determination of Peierls barrier~\cite{xu1998kink}. 
If the kink migration barrier is much lower than the screw dislocation migration barrier, the kink velocity can be used in place of the migration barrier in  kMC simulations~\cite{Cai2001kmc}.
All above allow for the effective parameterization of long dislocations for kMC simulations (e.g., those proposed by Cai and collaborators~\cite{Cai2001kmc,bulatov2006computer}) of long dislocations in complex materials.

%%%%%%%%%%%%%%%%%%%%%%%%%%%%
\section{Conclusion}

We have studied the finite-temperature core structures of  $\langle \mathbf{a}\rangle$ screw dislocations in HCP Ti, through a multi-scale framework.
First, we characterize atomic interactions in Ti based upon machine learning, Deep Potentials (DP), which reproduce the stable/metastable dislocation core structures found via quantum mechanical, DFT calculations. 
DP was employed in molecular dynamics (MD) simulations of screw dislocation core structure at finite temperatures.
MD provides the statistics of directional dislocation  core dissociation ($\unslant{\pi}$, P and B cores) and  directional core glide ($\unslant{\pi}$-, P- and B-glide). 
We found that the $\unslant{\pi}$ core is stable and the P core is metastable, consistent with  0~K DFT results, while the B core is metastable above 300~K. 
Contrary to common understanding, the glide direction need not  align with the core dissociation direction; e.g., $\unslant{\pi}$ core can glide on the P plane. 

The MD observations allow us to identify all important unit kinetic events associated with dislocation core motion. 
The events were categorized as either core transition  (change in the dissociation direction) or core glide events (unit displacement along a slip plane). 
These events were incorporated into  a kinetic model and that was parameterized through the MD data. 
The machine learning-based fitting procedure ensures that the frequency of each core structure and the translational and rotational diffusion coefficients produced by a kinetic Monte Carlo (kMC) simulation implementation of the model are consistent with  MD data.  
We found that  P core glide on the P plane event has the lowest core glide barrier, the transition between P and $\unslant{\pi}$ cores has the lowest barrier among all  core transition events, and the glide of the $\unslant{\pi}$ core (on any plane) is very difficult. 

With the parameters (barriers and frequencies) obtained by fitting, the proposed kMC simulation procedure is applicable to dislocation core dynamics at any temperature and  applied stress. 
The dislocation will undergo a random walk (diffusion) in the absence of an applied stress; long-time dislocation core trajectories provide anisotropic translational and rotational diffusion coefficients. 
The former indicates that dislocation motion on the P plane is  fastest and  motion on the B plane is  slowest. 
This implies that the $\unslant{\pi}$ core is difficult to rotate and is stable while rotation away from the B core is fast. 
Under an applied stress, dislocation motion occurs through a locking-unlocking process at low temperatures, consistent with  experimental observations. 
The locking behavior originates from the high energy barrier associated with $\unslant{\pi}$ core glide. 
Application of different stress states yields that the motion of $\langle\mathbf{a}\rangle$ screw core in Ti is anisotropic. 
The temperature dependence of this anisotropy is consistent with the (limited) MD predictions. 

This work demonstrates that the intrinsic dynamic behavior of dislocations cannot be described based upon studies of 0~K dislocation core structures alone.  
Rather, statistical examination of the finite-temperature core structures is essential to determine,  not only, the finite-temperature stability of different core structures, but also the kinetic properties of dislocation motion and core structure transitions. 
The kinetic model and parameters obtained in this work provide the necessary inputs for the higher-level approaches, such as  kMC simulation of  long dislocation line (for which kink nucleation and propagation are considered) and  discrete dislocation dynamics. 
While the present study focuses on screw dislocation motion in Ti, the method described here is applicable to all dislocation types, at any temperature and stress state, in both simple (cubic) and complex (non-cubic) crystalline materials.

%%%%%%%%%%%%%%%%%%%%%%%%%%%%%%%%%%%%%%%%%%%%%
\section*{Methods}
\subsection*{MD Simulations}
The simulations employ a Deep Potential (DP) trained using DFT results for perfect crystals and defects in the $\alpha$, $\beta$ and $\omega$ phases of Ti~\cite{wen_Ti}. This DP successfully reproduces the 0~K core structures of the $\langle \mathbf{a}\rangle$ screw dislocations in Ti as predicted by DFT~\cite{wen_Ti}.

The MD simulation cell geometry is shown in Fig.~\ref{fig:model}a. 
An HCP  $\alpha$-Ti single crystal cylinder is constructed such that the $[1\bar{2}10]$ is parallel to the cylinder axis; the  Cartesian coordinate system employed has $\mathbf{e}_x \parallel [10\bar{1}0]$, $\mathbf{e}_y \parallel [0001]$ and $\mathbf{e}_z \parallel [1\bar{2}10]$. 
The simulation cell is periodic along the $\mathbf{e}_z$-axis. 
An $\langle \mathbf{a}\rangle$ screw dislocation, with both Burgers vector and line direction parallel to  $\mathbf{e}_z$, is introduced in the center of the cylinder by displacing the atoms according to the anisotropic elasticity solution (lattice parameters and elastic constants for this potential as a function of temperature). 
The configuration is equilibrated at different temperatures in an $NVT$ ensemble. 
The interactions between atoms in Region I  (see Fig.~\ref{fig:model}a) are described by the DP for Ti. 
The atoms in Region II are described as an Einstein crystal, i.e., the atoms are tethered at the coordinates of the as-constructed (with anisotropic elastic displacements of atoms from their perfect crystal locations) configuration by harmonic springs, to avoid the free surface of Region II which will apply image force on the dislocation core. The interface between Region I and II  has negligible effect on the simulation results. Details can be found in the SI. 
The spring constant is determined as 
$3k_\txB T/\langle (\Delta\mathbf{r}_\text{atom})^2\rangle$, 
where $k_\txB$ is the Boltzmann constant, $T$ is the absolute temperature and $\langle (\Delta\mathbf{r}_\text{atom})^2\rangle$ is the mean squared displacement of atoms at the temperature $T$. 
The radius of Region I is $\sim160$~\AA, the width of Region II is  $\sim18$~\AA, and the dislocation line length is $\sim6$~\AA.
The whole system contains $\sim33,000$ atoms; i.e., large enough that Region II  and the interface between Region I and II have little influence on the random walk of the dislocation core about the center of Region I; see SI for details. 
The cylindrical sample containing a dislocation was equilibrated for 2~ns at each temperature. 
All  MD simulations were performed using {\sc lammps}~\cite{plimpton1995fast}. 

The dislocation configurations were thermally equilibrated at  temperatures  300-900~K (well below the HCP-BCC transition temperature, 1250~K). 
The atomic configuration was recorded every 50~fs. 
Energy minimization for 10-20 steps (by conjugate gradient with  line-search step size 0.01~{\AA}) was conducted for each of these atomic configurations to remove  thermal vibration in order to clearly visualize/analyze the atomic structure~\cite{bulatov2006computer}. Such energy minimization has little effect on the dislocation core distribution (for details, see SI).

\subsection*{Nye tensor parameters}
The Nye tensors are visualized with perfect crystals as references. 
Two key parameters govern the presentation of the Nye tensor plot are the cutoff distance for constructing a neighbor list and the maximum angle ($\Theta$) employed to identify matches between $\mathbf{p}$ and $\mathbf{q}$ vectors -- here, $\mathbf{p}$ and $\mathbf{q}$ denote the radial distance vectors between each atom and its neighbors  in the reference and current systems, respectively. 
In our investigation, we have chosen a cutoff distance that corresponds to $1.3$ times the equilibrium lattice constant at the temperature of interest. We  set $\Theta$ to $10^\circ$.

%%%%%%%%%%%
\subsection*{Prediction of event frequencies associated with dislocation dynamics}
In \textit{S1} of Sect.~\ref{sect:coredynamics}, we need an initial guess of the frequencies of all events at all temperatures based on the MD results. 
We treat the core transition events ($i\ne j$) and the core glide events ($i=j$)  differently. 

The frequency for a core transition event $i(\kappa)j$ ($i\ne j$), $\nu_{i(\kappa)j}$, is obtained from the MD results by 
\begin{equation}
\nu_{i(\kappa)j} 
= \frac{\mathcal{N}_{i(\kappa)j}}{\mathcal{N}_i\Delta t}
= \frac{\mathcal{P}_{i(\kappa)j}}{\Delta t}, 
\end{equation}
where $\Delta t$ is the time interval between recording atomic configurations, $\mathcal{N}_i$ is the number of configurations showing the $i$ core, $\mathcal{N}_{i(\kappa)j}$ is the number of the $i(\kappa)j$ events recorded, and $\mathcal{P}_{i(\kappa)j}$ is the conditional probability $\mathcal{P}(i\to j| i) = \mathcal{P}(i\to j)/\mathcal{P}(i) = \mathcal{N}_{i(\kappa)j}/\mathcal{N}_i$. 
The sampling interval $\Delta t$ is chosen small enough that few events are missed but not so small that thermal vibrations lead to incorrect event identification. 
We set $\Delta t = 0.1$~ps (i.e., close to the inverse Debye frequency for Ti). 

The frequency for a core glide event $i(\kappa)i$, $\nu_{i(\kappa)i}$, may also be extracted from the MD simulation atomic configurations, sampled with time interval $\Delta t$. 
Suppose that the number of the $i(\kappa)i$ events is $\mathcal{N}_{i(\kappa)i}$ and the displacement vector of the $m^{\text{th}}$ $i(\kappa)i$ event is $\mathbf{d}_{i(\kappa)i}(m)$. 
Then, the total time spent on the $i(\kappa)i$-glide is $\mathcal{N}_{i(\kappa)i} \Delta t$ and the glide distance of the $i(\kappa)i$ event  is $\mathbf{s}_\kappa \cdot \mathbf{d}_{i(\kappa)i}(m)$, where $\mathbf{s}_\kappa = \mathbf{n}_\kappa \times \mathbf{b}/|\mathbf{b}|$ ($\mathbf{n}_\kappa$ is the normal to the $\kappa$ plane and $\mathbf{b}$ is the screw dislocation Burgers vector). 
The frequency for the $i(\kappa)i$-glide event is thus obtained from 
\begin{equation}\label{eq:glide_frequency}
\nu_{i(\kappa)i} 
= \frac{1}{\mathcal{N}_{i(\kappa)i} \Delta t}
\cdot \frac{\mathbf{s}_\kappa}{2L_{i(\kappa)i}}
\cdot 
\sum_{m=1}^{\mathcal{N}_{i(\kappa)i}} 
\mathbf{d}_{i(\kappa)i}(m), 
\end{equation}
where $L_{i(\kappa)i}$ is the shortest glide distance, i.e., one period of Peierls barrier along the direction $\mathbf{s}_\kappa$. 
$\mathbf{s}_\txB$, $\mathbf{s}_\txP$ and $\mathbf{s}_{\unslant{\pi}}$ are the unit vectors in the $[10\bar{1}0]$, $[0001]$ and $[10\bar{1}2]$ directions, respectively. 
According to Figs.~\ref{fig:events}a-f, $L_{\txB(\txB)\txB} = \sqrt{3}a/2$, $L_{\txP(\txP)\txP} = c/2$,  $L_{\unslant{\pi}(\txP)\unslant{\pi}} = c$ and $L_{\unslant{\pi}(\unslant{\pi})\unslant{\pi}} = \sqrt{3a^2/4+c^2}$. 

\subsection*{Kinetic Monte Carlo simulations of dislocation dynamics}\label{KMC_simulation}
Here, we  focus on simulating the motion of a short, straight $\langle \mathbf{a}\rangle$ screw dislocation line in Ti. 
The dislocation line in a periodic box is shorter than the correlation length along the dislocation; in this sense, we are considering  dislocation core dynamics, rather than the dynamics of a long dislocation line. 

We construct a 2D lattice in the $\mathbf{e}_x$-$\mathbf{e}_y$ plane. 
The period of the 2D lattice is $a_x = \sqrt{3}a/2$ and $a_y = c$ in the $\mathbf{e}_x$ and  $\mathbf{e}_y$-directions (Fig.~\ref{fig:model}c). 
All  admissible events are listed in Fig.~\ref{fig:events}. 
The core state at a particular time is described by the core structure``$i$'' and location $(x,y)$. 
Some events involve glide in two opposite directions (i.e., corresponding to the ``$\pm$''  in the  ``Transition ($\kappa$)'' column in Fig.~\ref{fig:events}g. 
When these events occur, ``$+$''/``$-$'' is chosen randomly. 
The  kinetic Monte Carlo (kMC) algorithm is:
\begin{enumerate}[font=\itshape,label=\alph*:]
       \itemsep0em 
        \item
        Assume a dislocation core initially dissociated as $i=$ P, $\unslant{\pi}_1$, $\unslant{\pi}_2$ or B, and $(x,y)=(0,0)$.    
        \item
        Generate a list of admissible events $i(\kappa)j$ according to Fig.~\ref{fig:events}g and the corresponding frequencies $\{\nu_{i(\kappa)j}\}$ (extracted from the MD results). 
        Calculate the ``activity'': $\Lambda = \sum_{\kappa, j} \nu_{i(\kappa)j}$. 
        \item
        Randomly choose an event with  probability $\nu_{i(\kappa)j}/\Lambda$ (see \cite{lesar2013computational} for details). 
        \item
        Advance the kMC clock by $\Delta t_{\text{kmc}} = \Lambda^{-1} \ln(R^{-1})$, where $R$ is a random number in the range $(0,1]$.  
        \item
        Update the state, including the core structure $i$ and $(x,y)$, by the chosen event. 
        Return to \it{a}. 
\end{enumerate}

\subsection*{Optimization of event frequencies associated with dislocation dynamics}

The frequencies $\{\nu_{i(\kappa)j}\}$ are adjusted such that the MSD, the MSAD and the core probabilities $\{\mathcal{P}_i\}$ obtained from the kMC simulation are consistent with the MD results. 
The loss function used in this optimization is 
\begin{align}
&\mathcal{L}
= \lVert \mathbf{X}_\text{kmc}(\{\nu_{i(\kappa)j}\}) - \mathbf{X}_\text{md} \rVert^2
\qquad\text{and}
\nonumber\\
&\mathbf{X} 
\equiv (D_x^\txT, D_y^\txT, D^\txR_\txP, D^\txR_{\unslant{\pi}}, D^\txR_\txB, \mathcal{P}_\txP, \mathcal{P}_{\unslant{\pi}}, \mathcal{P}_\txB), 
\end{align}
where $\lVert \cdot \rVert$ denotes the $L^2$-norm, the subscripts ``kmc''/``md'' denote the quantities obtained from kMC/MD simulations and $\mathbf{X}_\text{kmc}$ is a function of $\{\nu_{i(\kappa)j}\}$. 
$\mathcal{L}$ is minimized with respect to $\{\nu_{i(\kappa)j}\}$.

\section*{Acknowledgements}
JH acknowledges support from the  Hong Kong Research Grants Council Early Career Scheme Grant CityU21213921 and Donation for Research Projects 9229061.
DJS gratefully acknowledge the support of the  Hong Kong Research Grants Council Collaborative Research Fund C1005-19G. 

\section*{Author contribution}
AL, JH and TW performed the research and analyzed the data. JH and DJS conceived and directed the project. AL, JH and DJS wrote, discussed and commented on the manuscript.

\section*{Competing interests}
The authors declare no competing interests.

\section*{Data availability}
The data that support the findings of this study are available from the corresponding author upon reasonable request.

\section*{Code availability}
The codes in this study are available from the authors upon reasonable request.

\bibliography{mybib}

\end{document}